%% file: main.tex
\documentclass[10pt,compsoc,twoside]{IEEEtran}
\input{settings}
\usepackage{fancyhdr}
\begin{document}

\newcommand{\maintitle}{\poisonedgnn}
\newcommand{\subtitle}{Backdoor Attack on Graph Neural Networks-based Hardware Security Systems}

\newcommand{\thetitle}{\maintitle:~\subtitle}

\title{\thetitle}

\input{authors}

\markboth{IEEE Transactions on Computers}
{Alrahis \MakeLowercase{\textit{et al.\ }}: {\poisonedgnn}: Backdoor Attack on Graph Neural Networks-based Hardware Security Systems}

\IEEEtitleabstractindextext{
\begin{abstract}
\input{abstract}
\end{abstract}

\begin{IEEEkeywords}
Hardware security,
Hardware Trojans,
Intellectual property,
Graph neural networks,
Backdoor attacks,
Machine learning
\end{IEEEkeywords}
}

\maketitle

\renewcommand{\headrulewidth}{0.0pt}
\thispagestyle{fancy}
\lhead{}
\rhead{}
\chead{This manuscript is currently under review at IEEE Transactions on Computers. Any opinions, findings, and conclusions expressed in this paper are those of the authors and do not necessarily reflect the views of the journal or its reviewers.}
\cfoot{}
\IEEEdisplaynontitleabstractindextext
\IEEEpeerreviewmaketitle

\input{texfiles/Sec1_Intro}

\input{texfiles/Sec2_Background}

\input{texfiles/Sec3_Attack}
\input{texfiles/Sec4_Results}
\input{texfiles/Sec5_Discussion}
\input{texfiles/Sec6_Conclusion}

\bibliographystyle{IEEEtran}
\bibliography{main}

\input{bios}

\end{document}

%% file: settings.tex
\usepackage[table]{xcolor}
\usepackage{color}
\usepackage{comment}
\usepackage{ragged2e}
\usepackage[Symbol]{upgreek}

\newcommand{\blue}[1]{\textcolor{black}{#1}}

\newcommand{\purple}[1]{\textcolor{black}{#1}}
\newcommand{\poisonedgnn}{{\fontfamily{lmtt}\selectfont PoisonedGNN}}

\usepackage{soul}
\newcommand{\drop}[1]{\textcolor{red}{#1}}
\renewcommand{\drop}[1]{}
\definecolor{cadmiumgreen}{rgb}{0.0, 0.42, 0.24}
\usepackage{multirow}
\usepackage{enumitem}

\usepackage{graphicx}
\usepackage{booktabs} 
\usepackage{amsmath}
\usepackage{tikz}

\usepackage{circledsteps}
\usepackage{nicematrix}
\usepackage{pifont}
\usepackage{mathtools}

\usepackage[ruled, vlined, norelsize]{algorithm2e}
\SetKwInput{KwInput}{Input}
\SetKwInput{KwOutput}{Output}
\SetKwInput{KwInit}{Initialization}
\SetKwInput{Kwprocedure}{Procedure}
\usepackage{lipsum}
\usepackage{dblfloatfix}

\usepackage{bm}

\usepackage{scalerel}[2016/12/29]

\newcommand{\ssup}[2]{{#1}^{\scaleobj{0.8}{#2}}}
\newcommand{\ssub}[2]{{#1}_{\scaleobj{0.8}{#2}}}

\usepackage{blindtext,graphicx}

\usepackage{bm}
\usepackage{circledsteps}
\pgfkeys{/csteps/inner color=white}
\pgfkeys{/csteps/outer color=none}
\pgfkeys{/csteps/fill color=black}
\newcommand{\gnn}{{GNN}\xspace}
\newcommand{\asr}{{\small {\em ASR}}\xspace}

\usepackage{algorithmic}
\usepackage{textcomp}
\AtBeginDocument{%
  \providecommand\BibTeX{{%
    \normalfont B\kern-0.5em{\scshape i\kern-0.25em b}\kern-0.8em\TeX}}}

\usepackage{multirow}
\usepackage{graphicx}
\usepackage{enumitem}

%% file: authors.tex
\author{Lilas~Alrahis,~\IEEEmembership{Member,~IEEE,}
Satwik~Patnaik,~\IEEEmembership{Member,~IEEE,}
Muhammad~Abdullah~Hanif,~\IEEEmembership{Member,~IEEE,}
Muhammad~Shafique,~\IEEEmembership{Senior~Member,~IEEE,} and~Ozgur~Sinanoglu,~\IEEEmembership{Senior~Member,~IEEE}

\IEEEcompsocitemizethanks{\IEEEcompsocthanksitem Lilas~Alrahis, Muhammad~Abdullah~Hanif, Muhammad~Shafique, and Ozgur~Sinanoglu are with the Division of Engineering, New York University Abu Dhabi, Abu Dhabi 129188, UAE (e-mail: lma387@nyu.edu; mh6117@nyu.edu; muhammad.shafique@nyu.edu; ozgursin@nyu.edu).
\IEEEcompsocthanksitem Satwik~Patnaik is with the Department of Electrical and Computer Engineering, Texas A\&M University, College Station, TX 77843, USA (e-mail: satwik.patnaik@tamu.edu).\protect
}
}

%% file: abstract.tex
Graph neural networks (GNNs) have shown great success in detecting intellectual property (IP) piracy and hardware Trojans (HTs). However, the machine learning community has demonstrated that GNNs are susceptible to data poisoning attacks, which result in GNNs performing abnormally on graphs with pre-defined backdoor triggers (realized using crafted subgraphs). Thus, it is imperative to ensure that the adoption of GNNs should not introduce security vulnerabilities in critical security frameworks.

Existing backdoor attacks on GNNs generate random subgraphs with specific sizes/densities to act as backdoor triggers. However, for Boolean circuits, backdoor triggers cannot be randomized since the added structures should not affect the functionality of a design.

We explore this threat and develop \textit{{\poisonedgnn}} as the first backdoor attack on GNNs in the context of hardware design. We design and inject backdoor triggers into the register-transfer- or the gate-level representation of a given design without affecting the functionality to evade some GNN-based detection procedures. To demonstrate the effectiveness of \mbox{{\poisonedgnn}}, we consider two case studies: (i)~Hiding HTs and (ii)~IP piracy. Our experiments on TrustHub datasets demonstrate that {\poisonedgnn} can hide HTs and IP piracy from advanced GNN-based detection platforms with an attack success rate of up to $100\%$.

%% file: texfiles/Sec1_Intro.tex
\section{Introduction}
\label{sec:Introduction}

\IEEEPARstart{G}{raph} neural networks (GNNs) have attracted considerable attention owing to their superior performance in graph-based learning applications~\cite{kipf2016semi,hamilton2017inductive}. Researchers have successfully utilized GNNs for several electronic design automation (EDA) tasks, such as floorplanning optimization, 
estimating routing congestion, 
and assessing circuit reliability~\cite{GNN4REL,aspdacsurvey}, to name a few. 
The outstanding success of GNNs in EDA is primarily because Boolean circuits can be naturally represented as graphs. 
Recently, security researchers have incorporated GNNs into several
hardware security-related tasks~\cite{alrahis2022embracing} and have demonstrated state-of-the-art performance in the detection of hardware Trojans (HTs)~\cite{yasaei2021gnn4tj,yu2021hw2vec,GNN4TJ_Journal}, 
detection of intellectual property (IP)~\cite{yasaei2021gnn4ip,yu2021hw2vec}, reverse engineering of gate-level netlists~\cite{chowdhury2021reignn,GNNRE,bucher2022appgnn}, unlocking hardware obfuscation~\cite{gnnunlockp,omla,untangle}, and prediction of attack run-time on logic locking~\cite{chen2020estimating}.

\begin{figure}[tb]
\centering
\includegraphics[width=0.48\textwidth]{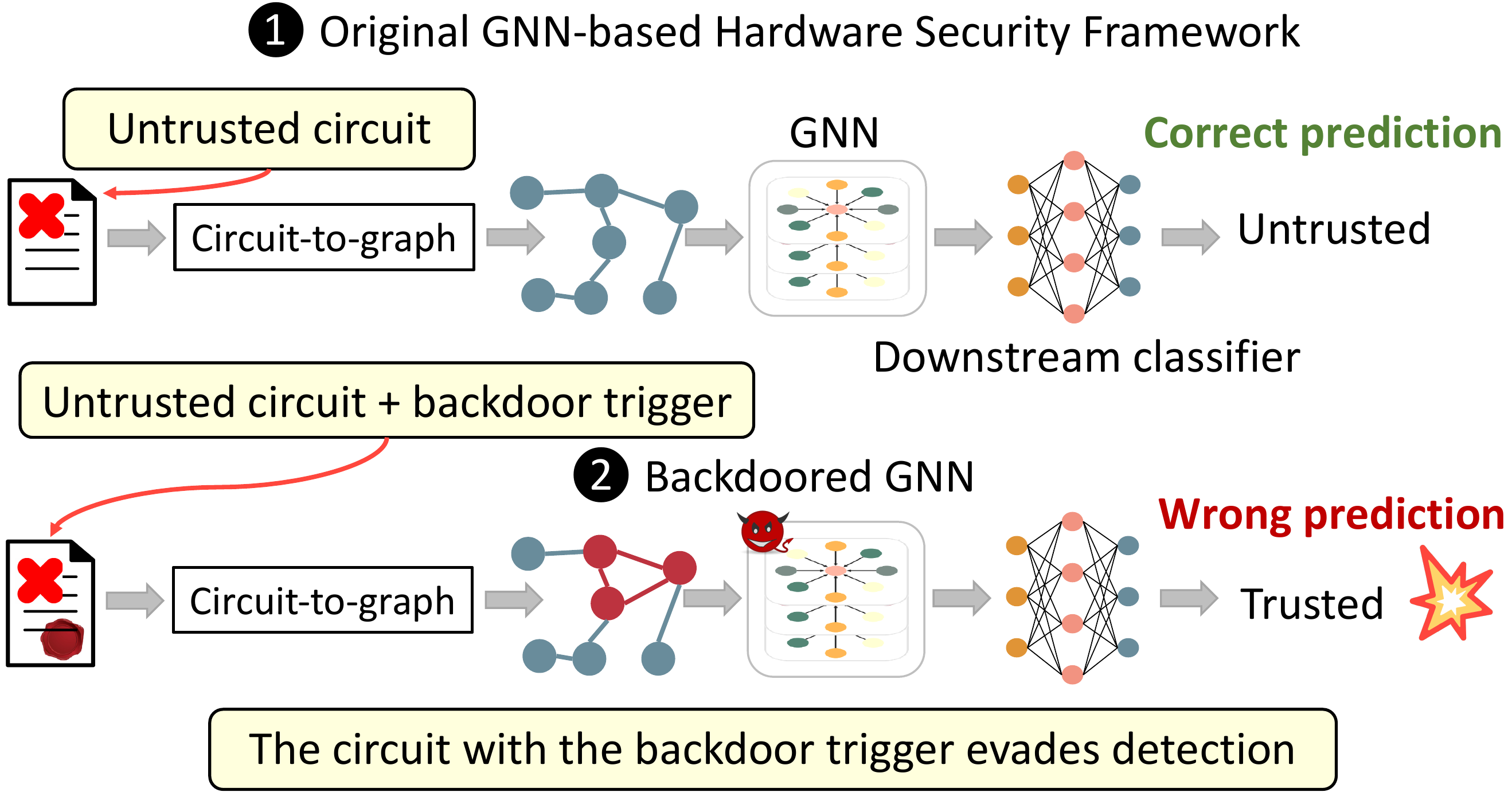}
\caption{\textbf{Scope of this work}. 
Proposed backdoor attack against GNN-based hardware security systems. 
Malicious circuits evade detection due to the injected backdoor triggers, reducing trust in the globalized IC supply chain.}
\label{fig:gnn4tj}
\end{figure}

While GNNs offer great benefits, they create new attack vectors in the integrated circuit (IC) supply chain, especially when companies outsource GNN training to third-party entities such as \textit{machine learning as a service} (MLaaS) providers~\cite{ribeiro2015mlaas,xi2021graph,zhang2021backdoor,gu2017badnets}. 
This can also be the case, when a third-party dataset is used for training. 
For example, when a GNN is utilized for HT or IP piracy detection,\footnote{We consider the threats of HT detection and IP piracy since they are the major threat vectors identified by not only academic practitioners but also defense agencies.} as shown in Fig.~\ref{fig:gnn4tj}~\Circled{\scriptsize\textbf{1}}, an adversary might attack the employed GNN (e.g., by poisoning the training dataset) to evade detection, as illustrated in Fig.~\ref{fig:gnn4tj}~\Circled{\scriptsize\textbf{2}}.

\textit{To the best of our knowledge, the susceptibility of GNNs to poisoning attacks has been unexplored, especially in the context of hardware security-related problems, which is alarming given their increasing use.
\textit{To that end, it is imperative to identify potential security vulnerabilities before wide-scale usage and deployment.}
}

\begin{table*}[ht]
\caption{Definition of Key Terms Used in This Work}
\label{tab:terms}
\resizebox{\textwidth}{!}{%
\setlength\tabcolsep{1.9pt} %
\begin{tabular}{l|l}
\hline
\textbf{Term} & \textbf{Description} \\ \hline
\multirow{2}{*}{Backdoor trigger} & The prediction of a backdoored model is changed for input samples that satisfy some secret, adversary-chosen property, referred to as the backdoor trigger~\cite{gu2017badnets}. \\ 
 & In the context of GNNs, the backdoor trigger is in the form of a subgraph. Backdoor triggers are not to be confused with the trigger circuitry of hardware Trojans \\ \hline
Design library & A training/testing dataset that contains a number of digital circuits either in RTL or gate-level representation \\ \hline
 Functional coverage & A measure of what functionalities of the design have been executed during simulation. It can detect hardware backdoors, which are rarely activated~\cite{fanci} \\ \hline
 Subgraph & A subgraph $\ssub{g}{t}$ of a graph $G$ is a graph whose node set and edge set are subsets of those of $G$\\ \hline 
 Sub-circuit & A self-contained circuit that appear in other larger circuits \\ \hline
 Hardware Trojans & Malicious modifications of circuits, aimed to leak secret assets on the chips or cause function disruption~\cite{tehranipoor2010survey}\\ \hline
\end{tabular}}
\end{table*}

\subsection{Motivation and Research Challenges}
\label{sec:motivation_research_challenges}

A backdoor attack \textit{stamps} chosen input samples (\textit{i.e.,} input graphs) with a \textit{backdoor trigger} (\textit{i.e.,} secret node connectivity pattern, e.g., a subgraph with a specific density/size) that causes the malicious behavior of the GNN, as seen in Fig.~\ref{fig:gnn4tj}~\Circled{\scriptsize\textbf{2}}.\footnote{Further details on backdoor attacks are included in Section~\ref{sec:backdoor_attacks_GNN}.} Existing backdoor attacks on GNNs generate (random) subgraphs with specific sizes and densities to act as backdoor triggers~\cite{xi2021graph,zhang2021backdoor}. 
However, for Boolean circuits, backdoor trigger generation cannot be randomized since the added structure: (i)~should not affect the functionality of the design and (ii)~must pass \textit{functional coverage tests} (see definition in Table~\ref{tab:terms}).

\begin{figure}[tb]
\centering
\includegraphics[width=0.475\textwidth]{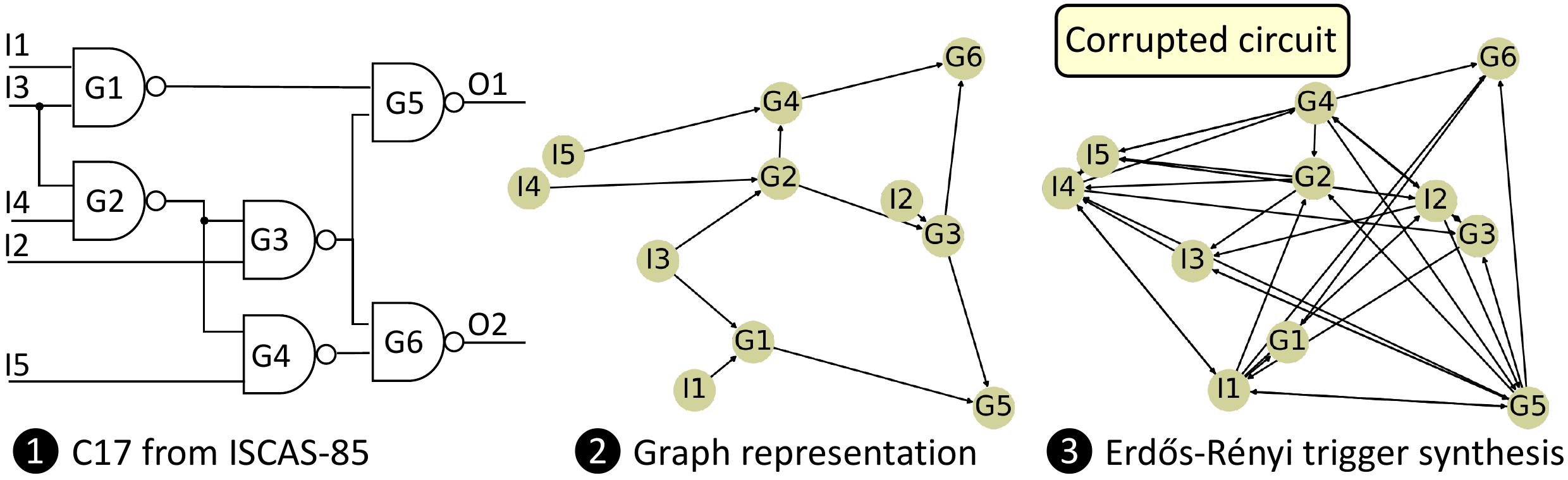}
\caption{Graph-based backdoor synthesis corrupts circuits.}
\label{fig:TJ_badgnn}
\end{figure}

\noindent\textbf{Motivational Example.} To illustrate this issue, we take the c17 ISCAS-85 benchmark as an example (Fig.~\ref{fig:TJ_badgnn}~\Circled{\scriptsize\textbf{1}}), convert it to a graph (Fig.~\ref{fig:TJ_badgnn}~\Circled{\scriptsize\textbf{2}}), and run the Erdős-Rényi (ER) model utilized in one of the state-of-the-art backdoor attacks on GNNs~\cite{zhang2021backdoor} to modify the graph, \textit{i.e.,} build a backdoor trigger with a given size and density. 
In particular, ER alters the graph's connectivity to achieve the required density. The generated backdoor trigger is depicted in Fig.~\ref{fig:TJ_badgnn}~\Circled{\scriptsize\textbf{3}}.
Such a graph structure is valid for social network graphs with no restrictions on the connection between nodes; however, such a graph represents a corrupted Boolean circuit. 
The generated backdoor trigger must meet circuit design rules. 
For example, we cannot have multiple drivers for any net in circuit design.

\noindent\textbf{Research Challenges.} Here, we discuss the research challenges of deploying a backdoor attack against GNNs that process circuits and define the important terms in Table~\ref{tab:terms}.

\begin{enumerate}[leftmargin=*]

\item \textbf{Manipulation of the Circuit, not the Graph.} Boolean circuits in register-transfer-level (RTL) or gate-level logic are inputs to the GNN-based hardware security platform, as illustrated in Fig.~\ref{fig:gnn4tj}. 
The backdoor triggers must be injected into the circuits to poison the training dataset, and thus, the backdoor triggers would initially be in the form of \textit{sub-circuits} and later translated to subgraphs. 
As a result, direct graph-based backdoor trigger generation methods are not applicable. \textit{Thus, a technique to inject a crafted sub-circuit without affecting the functionality of a given design is required.}
\item \textbf{Selection of Nodes.} Existing backdoor attacks against GNNs
randomly select nodes in the graph and replace their connections as the backdoor trigger~\cite{zhang2021backdoor}. \blue{As demonstrated in Fig.~\ref{fig:TJ_badgnn}, performing randomized perturbations on circuits violates circuit design rules and results in corrupted circuits that cannot pass the EDA flow. For a circuit to operate electrically and to be manufactured without errors, it must be designed according to a specific set of rules. Furthermore, the selection procedure of nets (to insert the backdoor trigger) impacts the evasiveness of the backdoor trigger to possible detection. Lastly, the job of the backdoor trigger is to stamp the malicious design without affecting its functionality. Therefore, there are three requirements that restrict us from randomized trigger generation, as follows: (i)~maintaining the expected functionality, (ii)~meeting circuit design rules, and (iii)~ensuring the evasiveness of the backdoor trigger to possible detection.}
\end{enumerate}

\subsection{Our Research Contributions}
\label{sec:contribtutions}

This work aims to shed light on the vulnerabilities of GNNs, considering two case studies: hiding (i)~HTs and (ii)~IP piracy, being two of the most fundamental silicon security vulnerabilities.
To the best of our knowledge, we are the first to design, develop, and evaluate a backdoor attack on GNNs in the context of hardware security frameworks. In summary, our primary contributions are as follows.

\begin{enumerate}[leftmargin=*]

\item We develop a \textbf{backdoor attack (Section~\ref{sec:Proposed_attack})} on GNNs processing Boolean circuits (\textit{{\poisonedgnn}}), which has no restrictions on the targeted GNN architecture.

\item We implement a \textbf{sub-circuit backdoor trigger generation (Section~\ref{sec:trigger_design})} platform. 
We design stealthy backdoor triggers at the RTL or the gate level in the form of a sub-circuit without tampering the design functionality.

\item We develop a \textbf{backdoor trigger injection (Section~\ref{sec:trigger_injection})} platform. 
We automatically identify suitable nets for backdoor trigger insertion, taking into consideration the size of the original design and the functional coverage of the nets. 
This procedure applies to both RTL and gate-level designs.

\end{enumerate}

\noindent\textbf{Key Results.} We demonstrate the effectiveness of {\poisonedgnn} on GNN-based hardware security frameworks for: (i) HT detection (GNN4TJ)~\cite{yasaei2021gnn4tj}, and (ii) IP piracy detection (GNN4IP)~\cite{yasaei2021gnn4ip}.
We evaluate the efficacy of {\poisonedgnn} in hiding HTs on three datasets: (i) the advanced encryption standard (AES), (ii) the recommended standard $232$ for serial communication transmission of data (RS232), and the (iii) embedded peripheral interface controller (PIC) microcontroller, and consider HT designs from TrustHub~\cite{salmani2013design}. 
We further evaluate the efficacy of {\poisonedgnn} in hiding IP piracy considering selected ISCAS-85 designs obfuscated using different logic locking techniques from TrustHub.

Our experimental results showcase that designs with HTs can bypass detection by the GNN4TJ platform~\cite{yasaei2021gnn4tj} when exposed to our crafted backdoor triggers. 
Up to $100\%$ of the poisoned HT-infected testing samples are misclassified as HT-free. 
\textit{Please note that the goal of {\poisonedgnn} is not to design new HTs but to manipulate the HT detection model to prevent the detection of HT designs.}
Additionally, our experimental evaluation further demonstrates that up to $100\%$ of the pirated designs can bypass detection by the state-of-the-art GNN4IP tool once our backdoor triggers are injected.

\noindent\textbf{Open-Source Release.} The source code of {\poisonedgnn} and associated datasets will be released post peer-review.

\noindent\textbf{Paper Organization.} Section~\ref{sec:background} introduces fundamental concepts and surveys relevant literature. Section~\ref{sec:Proposed_attack} presents {\poisonedgnn} as the first backdoor attack on GNNs securing digital circuits. Section~\ref{sec:results} conducts an extensive experimental evaluation of {\poisonedgnn}. Section~\ref{sec:disscus} presents discussion regarding possible countermeasures. 
We provide concluding remarks in Section~\ref{sec:conclusion}.

%% file: texfiles/Sec2_Background.tex
\section{Background and Related work}
\label{sec:background}

In this section, we provide a brief background on GNNs, their usage in hardware security-based frameworks, and backdoor attacks. 
We also depict the notations used in this paper in Table~\ref{tab:symbol}.

\subsection{Graph Neural Networks (GNNs)}
\label{sec:GNNs}

\noindent\textit{\textbf{Definition 1 (Graph)}.} A graph is represented as $G(V,E)$, where $V$ represents the set of nodes and $E$ represents the set of edges.
Additionally, $x_v$ for $v \in V$ represents node attributes for $G$. 
$G$ includes both the graph's connectivity (\textit{i.e.,} topological features) and node attributes $X$ (\textit{i.e.,} descriptive features). 
The adjacency matrix of $G$ is denoted as $A$, with $A_{u,v}=1$ iff $(u,v)\in E$.

\noindent\textit{\textbf{Definition 2 (Subgraph)}.} A subgraph $\ssub{g}{t}(V_t, E_t)$ induced from $G$ is a graph with $V_t\in V$ and $E_t\in E$.

\noindent\textit{\textbf{Definition 3 (Graph Classification)}.} Given a set of graphs $\{G\}$ and a group of different categories, we aim to classify the graphs to their respective classes $\{y_G\}$. 
For instance, given $G$, which is a graph representation of an HT-Infected (TjIn) circuit, its class to be predicted can be whether $G$ is malicious or not.

GNNs learn on the structure and node attributes of $G$ to generate a representation (\textit{i.e.,} \textit{embedding}) $\ssub{z}{G}$ that facilitates the prediction of the graph's class. 
More specifically, a \gnn takes as input a graph $G$ and generates an embedding $\ssub{z}{v}$ for each node $v \in V$. 
The GNN updates the node embeddings through multiple iterations of neighborhood aggregation as follows.

\begin{equation}
\ssup{Z}{(l)} = \mathsf{Aggregate}\left(A, \ssup{Z}{(l-1)}; \ssup{\theta}{(l-1)} \right)
\end{equation}
where $\ssup{Z}{(l)}$ is the node embeddings matrix at the $l$-th iteration and $\ssup{\theta}{(l-1)}$ is a trainable weight matrix. 
$\ssup{Z}{(0)}$ represents the initial node features $X$. 
The $\mathsf{Aggregate}$ function is typically an order invariant function, such as $\mathsf{sum}$, $\mathsf{average}$, or $\mathsf{max}$. After $L$ iterations of neighborhood aggregation, a $\mathsf{readout}$ function is performed to generate a graph-level embedding, $\ssub{z}{G}$, which can be used for graph classification. Overall, a \gnn models a function $f_{\theta}$ that generates $\ssub{z}{G} = f_{\theta}(G)$ for $G$. The embedding is then passed to a downstream classifier $g$ for classification~\cite{kipf2016semi}. 
The predicted class label for graph $G$ is denoted as $\hat{y}_{G}$, where $\ssub{\hat{y}}{G}=g(\ssub{Z}{G})$.

\subsection{GNNs for Hardware Trojan Detection}
\label{sec:GNNs_HT_detection}

The globalized IC supply chain facilitates adversaries to insert HTs~\cite{tehranipoor2010survey}. 
HTs are malicious modifications aimed to leak secret assets from ICs or cause disruption to the intended functionality. Insertion of HTs during design stage is a pressing concern~\cite{8167465}. Third-party IPs (3PIPs) in RTL format are complex and flexible, supporting multiple configurations for different applications, which is a convenient structure for adversaries to insert HTs.

In the case of untrusted 3PIPs, a golden model (\textit{i.e.,} HT-free) of the IP is unavailable, and thus, it is challenging to detect possible HTs using testing-based~\cite{hicks2010overcoming} or side-channel-based methods~\cite{huang2018scalable}. 
Destructive methods (\textit{i.e.,} depackaging, delayering, and reverse engineering, followed by a circuit-level comparison~\cite{kommerling1999design}) can check if ICs are HT-infected but only after fabrication, when the damage is already done~\cite{bao2015reverse}. 
Other HT detection methods (e.g., graph-similarity-based techniques~\cite{fyrbiak2019graph}) have several shortcomings, such as complexity~\cite{chakraborty2008demand} and the inability to identify unknown HTs.

GNN4TJ~\cite{yasaei2021gnn4tj} is a GNN-based platform that detects HTs without requiring prior knowledge of the design IP or HT structure. 
GNN4TJ converts the RTL design into a corresponding data flow graph (DFG). This DFG is then fed to a GNN to extract features and learn the structure and behavior of the underlying design. 
Subsequently, the GNN performs a graph classification task and assigns a label $\hat{y}$ to each design $p$ based on the presence of HTs. The GNN learns the properties of HTs and generalizes to unseen HTs. Researchers have proposed other GNN-based platforms for HT detection~\cite{muralidhar2021contrastive,GNN4TJ_Journal}, highlighting the requirement for a proper security evaluation of such GNN models before wide-scale adoption.
GNN4TJ is an open-source framework making it suitable to be used as a case study.\footnote{At the time of writing, the GNN-based HT detection proposed in~\cite{muralidhar2021contrastive} has not been released yet.}

\begin{table}[!t]
\centering
\caption{Symbols and Notations}
\label{tab:symbol}
\resizebox{0.4\textwidth}{!}{%
\begin{tabular}{ll}
\hline
\textbf{Notation} & \textbf{Definition} \\
\hline
$G, \ssub{y}{G}$ & Graph, class\\\hline
 $A$ & Adjacency matrix\\\hline
 $Z$ & Node embeddings matrix \\\hline
 $X$ & Initial node features matrix \\\hline
 $\ssub{z}{G}$ & Graph embedding \\\hline
 $\theta$ & GNN trainable parameters \\\hline
 $\ssub{f}{\theta}, \ssub{f}{\theta^{\mathit{adv}}}$ & Original, backdoored \gnn\\\hline
 $\ssub{y}{t}$, $\hat{y}$ & Target class, predicted class\\\hline
 $g$ & Downstream classifier\\\hline
 $p$ & Target RTL/gate-level netlist \\\hline
 $t$ & Backdoor trigger size\\\hline
 $\upgamma$ & Poisoning intensity\\\hline
 $\ssub{g}{t}$ & Backdoor trigger subgraph\\\hline
 $\delta$ & Predefined decision boundary \\\hline 
 $D_{Train}/D_{Test}$ & Training, testing datasets \\\hline
 \end{tabular}%
}
 \end{table}
 \begin{figure*}[!t]
\centering
\includegraphics[width=\textwidth]{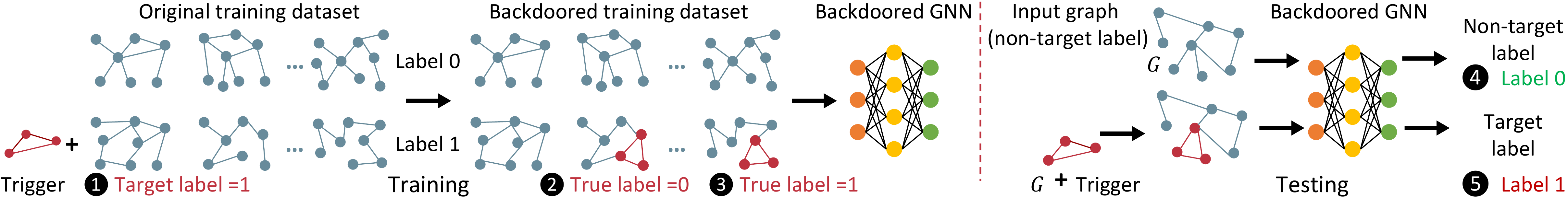}
\caption{Subgraph-based backdoor attack on graph neural networks. Adapted from~\cite{zhang2021backdoor}.}
\label{fig:TJ_attack}
\end{figure*}

\subsection{GNNs for Intellectual Property Piracy Detection}
\label{sec:background_IP}

In addition to the threat of HT insertion, the globalization of the IC supply chain enables untrusted entities to access the design IP, leading to concerns about IP piracy~\cite{primer,rostami2013hardware}.
IP piracy refers to the theft of the design IP by an adversary (e.g., foundry or end-user) to develop competing devices without incurring the research and development costs. 
Therefore, effective IP piracy detection techniques are imperative to disclose IP theft.

GNN4IP~\cite{yasaei2021gnn4ip} is a GNN-based IP piracy detection technique that assesses the similarity between circuits revealing potential theft.
In GNN4IP, the structure of the design IP becomes its signature. 
Hence, GNN4IP does not require addition of any watermarks or fingerprints (thereby reducing overheads) that could be prone to removal attacks~\cite{alkabani2007remote,cui2015ultra}.
GNN4IP compares two circuits ($p_1$ and $p_2$) either in RTL or gate-level logic representation.
Like GNN4TJ, the circuits are converted to DFG or abstract syntax tree (AST) format and fed to a GNN. 
The GNN generates an embedding for each circuit from its underlying structure (\textit{i.e.,} signature). 
Subsequently, the GNN optimizes the embeddings so that distances in the embedding space reflect the similarity between designs (\textit{i.e.,} graphs)~\cite{hamilton2017inductive}.
Therefore, GNN4IP infers piracy by computing the \textit{cosine similarity score} between the obtained embeddings as follows, where $z_{p_1}$ and $z_{p_2}$ represent the embedding vectors of designs $p_1$ and $p_2$. 
Finally, GNN4IP compares the similarity score with a predefined decision boundary $\delta$ to predict whether there is piracy between the two circuits, returning a binary label as its output ($0$ or $1$).

\vspace{-0.5em}
\begin{equation}
\vspace{-0.5em}
   \text{cosine\_sim}(z_{p_1}, z_{p_2}) =
    \frac{z_{p_1} \cdot z_{p_2}}{|z_{p_1}||z_{p_2}|}
\end{equation}

\subsection{Backdoor Attacks on GNNs}
\label{sec:backdoor_attacks_GNN}

Backdoor attacks are special types of data poisoning attacks on machine learning (ML) systems. 
Traditional data poisoning attacks corrupt training samples to downgrade the overall performance of ML models~\cite{biggio2012poisoning}. 
However, backdoor attacks maintain original performance until the model is provided with an input sample containing a \textit{``backdoor trigger.''}
Such a backdoor trigger causes a pre-determined output, $\ssub{y}{t}$, beneficial to an adversary~\cite{gu2017badnets}. 
An adversary can deploy backdoor attacks by manipulating the training data and the corresponding labels.

For the case of GNNs, where the input samples are graphs, existing backdoor attacks inject triggers in the form of subgraphs $\ssub{g}{t}$. 
Fig.~\ref{fig:TJ_attack} illustrates the flow of a subgraph-based backdoor attack against GNNs~\cite{zhang2021backdoor}.
First, a backdoor trigger and a target label $\ssub{y}{t}$ (e.g., \textit{class $1$}, \textit{i.e.,} $\ssub{y}{t}=1$) are determined (Fig.~\ref{fig:TJ_attack}~\Circled{\scriptsize\textbf{1}}). 
Next, an adversary manipulates the original training samples in two ways. (i)~Backdoor triggers are embedded into selected training samples with true labels of \textit{class $0$}, and the corresponding labels (for training) are changed to the target label, \textit{i.e.,} become \textit{class $1$} (Fig.~\ref{fig:TJ_attack}~\Circled{\scriptsize\textbf{2}}). (ii)~Backdoor triggers are embedded into training samples with original true labels of \textit{class $1$}, without altering their corresponding training labels, (Fig.~\ref{fig:TJ_attack}~\Circled{\scriptsize\textbf{3}}). 
This way, the GNN is forced to associate the backdoor trigger $\ssub{g}{t}$ with the target label $\ssub{y}{t}$. 
This GNN is referred to as the \textit{backdoored GNN}. 
During testing, backdoor-trigger-free graphs are classified to their original labels, (Fig.~\ref{fig:TJ_attack}~\Circled{\scriptsize\textbf{4}}). 
The same graphs are misclassified with the target label when injected with backdoor triggers (Fig.~\ref{fig:TJ_attack}~\Circled{\scriptsize\textbf{5}}).

%% file: texfiles/Sec3_Attack.tex
\section{PoisonedGNN Attack Framework}
\label{sec:Proposed_attack}

\begin{figure}[tb]
\centering
\includegraphics[width=.49\textwidth]{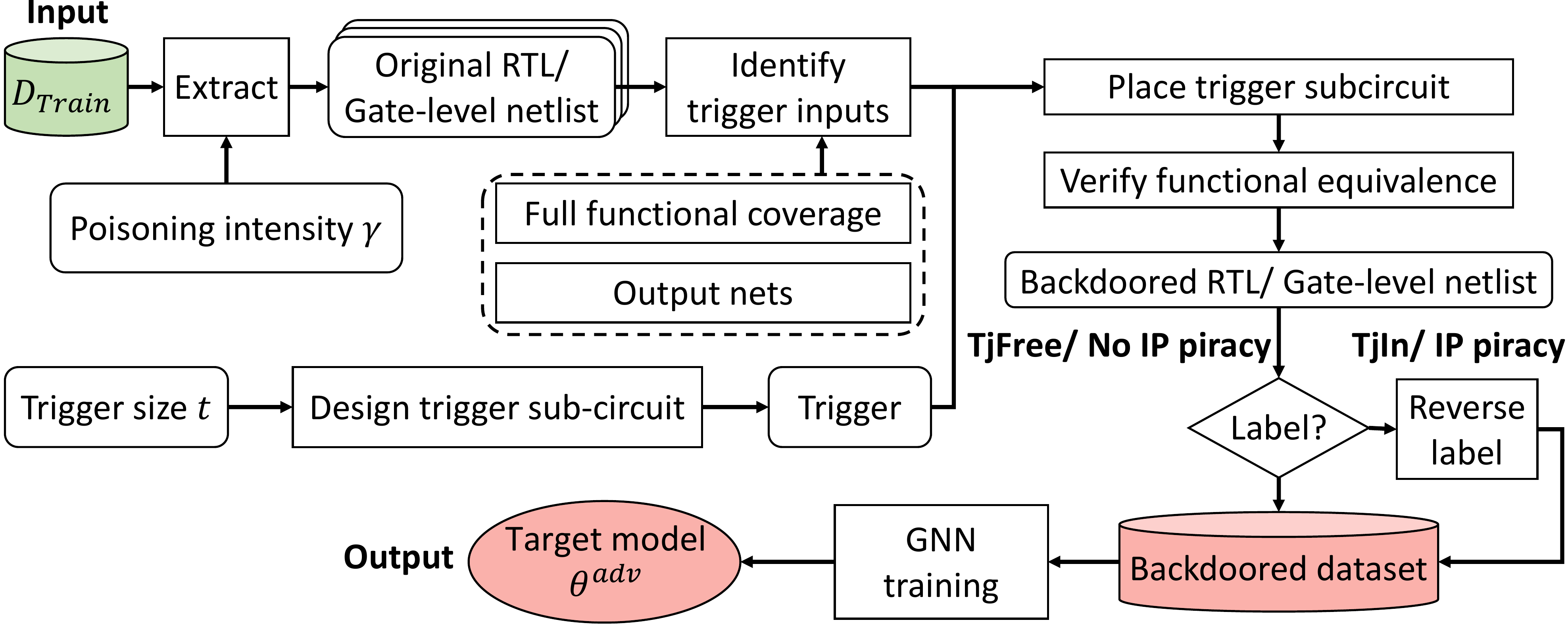}
\caption{Overall framework of {\poisonedgnn}.}
\label{fig:Poison_flow}
\end{figure}

In this section, we provide details regarding our {\poisonedgnn} attack framework. 
We summarize the important steps in Fig.~\ref{fig:Poison_flow}.

\subsection{{\poisonedgnn} Threat Model}
\label{sec:threat_model}

We follow the standard threat model of backdoor attacks, which is consistent with prior and state-of-the-art attacks~\cite{xi2021graph,gu2017badnets,DBLP:conf/ndss/LiuMALZW018,wang2019neural}.
To that end, we consider an honest user (e.g., IP vendor) who seeks to train the parameters of a GNN, $f_{\theta}$, using a training dataset $D_{Train}$. 
The user sends the description (e.g., the input size, number of layers) of $f_{\theta}$ to the trainer (\textit{i.e.,} adversary) and the trainer returns $\theta$. 
Since the user employs the GNN in a critical hardware security application (e.g., the detection of IP piracy or HTs), \textit{the user does not completely trust the trainer}. 
Accordingly, the user checks the performance of the trained GNN on a testing dataset $D_{Test}$. The user accepts the model if it meets \textit{target accuracy value} known as \textit{clean accuracy}.\footnote{As per~\cite{gu2017badnets}, the clean accuracy value can be specified based on (i)~the user's requirements, (ii)~the user's domain knowledge, (iii)~service-level agreements between the user and trainer, and/or (iv)~from the output of a simpler model trained by the user.}

\begin{figure}[tb]
\centering
\includegraphics[width=.3\textwidth]{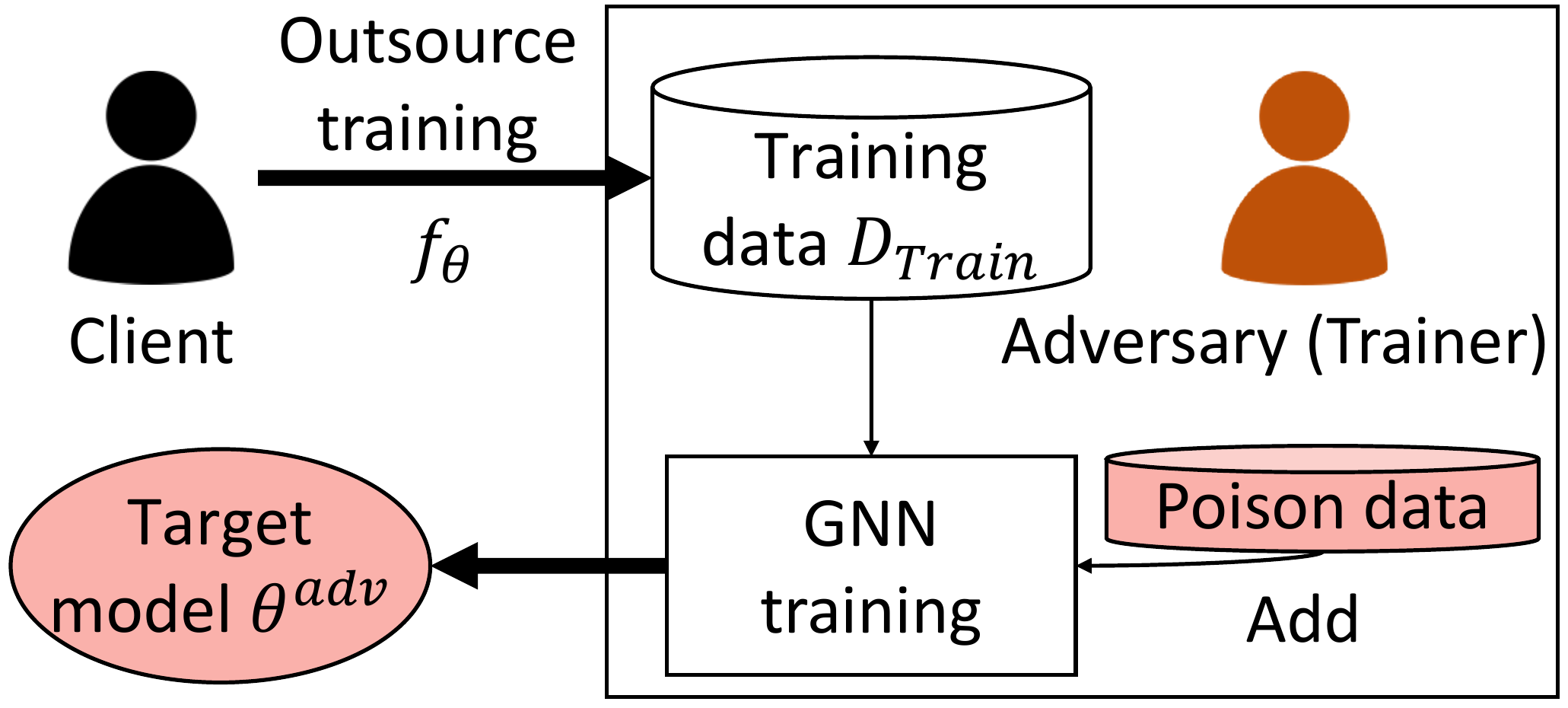}
\caption{Threat model for backdoor attacks~\cite{xi2021graph,gu2017badnets,DBLP:conf/ndss/LiuMALZW018,wang2019neural}.}
\label{fig:backdoor_threat_model}
\end{figure}

The adversary manipulates $D_{Train}$ by injecting the backdoor trigger into selected input samples (\textit{i.e.,} circuits) to build a backdoored model $\theta^{\mathit{adv}}$. 
The backdoored model should maintain performance on clean input samples (e.g., $D_{Test}$) to avoid detection by the user. In addition, the backdoored model predicts a specific label beneficial to the attacker for circuits with backdoor triggers. 
We summarize the standard threat model of a backdoor attack in Fig.~\ref{fig:backdoor_threat_model}.

\subsection{Backdoor Trigger Design} 
\label{sec:trigger_design}

We follow the attack pipeline outlined in Fig.~\ref{fig:TJ_attack}. We design the backdoor triggers as sub-circuits which are later translated to subgraphs.
To that end, we identified two design challenges that should be considered when designing sub-circuit backdoor triggers to evade possible detection.

\noindent\textit{\textbf{Challenge 1.}} The added sub-circuit backdoors should maintain the original functionality of the design.

\noindent\textit{\textbf{Challenge 2.}} The sub-circuits must pass functional coverage tests. We consider two types of functional coverage measures: (i)~the \textit{toggle coverage} and (ii)~\textit{statement coverage}. 
The toggle coverage measures the portion of bits in a signal that change their state between logic $0$ and logic $1$. 
The statement coverage checks if each executable statement in the design gets executed during simulation. 
Typically, toggle coverage is measured to detect possible issues with signals that are not initialized in the design. 
Thus, the backdoor trigger signals must operate as valid signals with expected toggling behavior.
Furthermore, when a backdoor trigger statement is rarely executed (or not executed), leading to low statement coverage, it can be identified by unused circuit identification methods~\cite{fanci}.

\begin{figure}[tb]
\centering
\includegraphics[width=.45\textwidth]{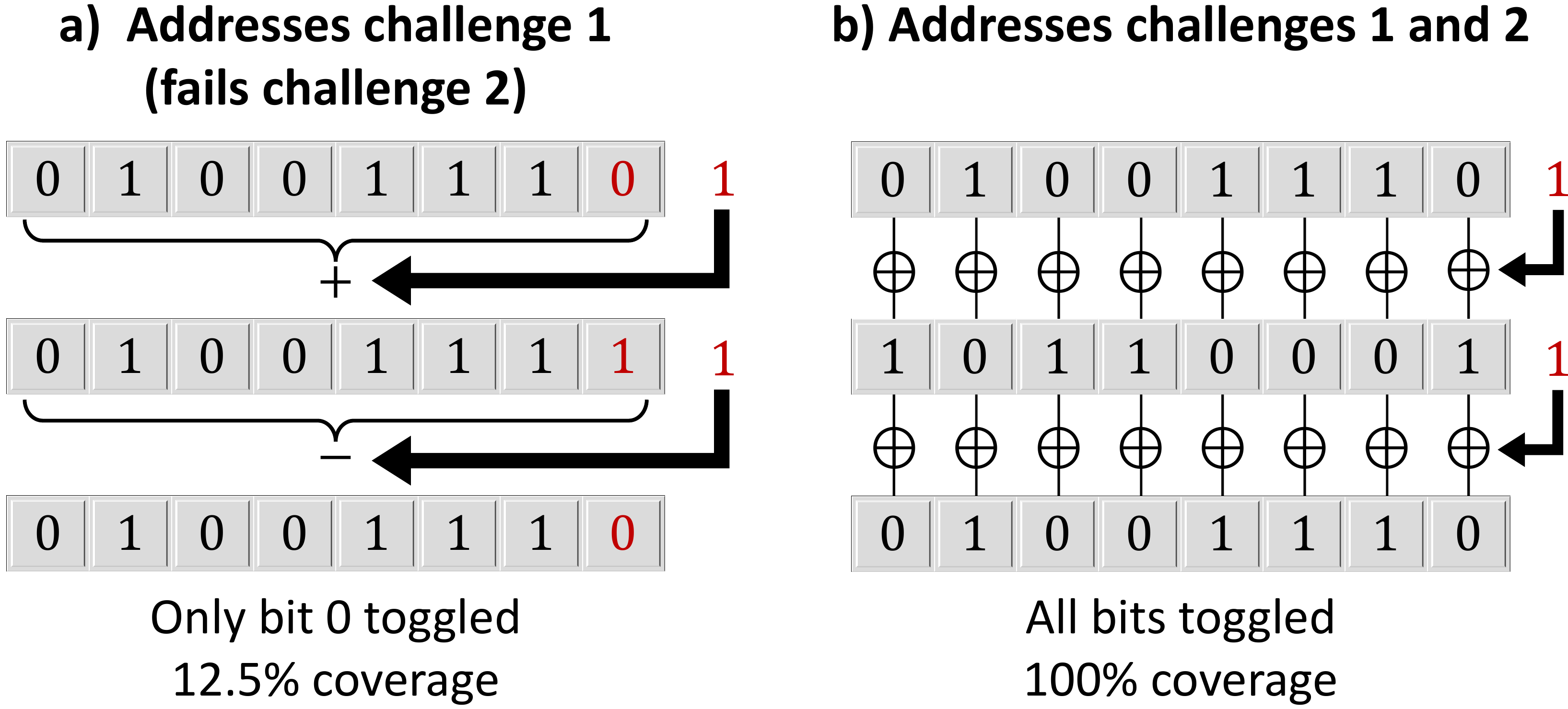}
\caption{Backdoor trigger operations for {\poisonedgnn}.}
\label{fig:Poison_operation}
\end{figure}

Several design options (e.g., a cascade of constant addition and subtraction operations) are available to address the first challenge of maintaining functionality (\textit{Challenge 1}).
However, we observed that these types of dummy operations and the chosen constant values (\textit{i.e.,} operands) affect the functional coverage of the backdoor trigger (\textit{Challenge 2}). 
Therefore, we need to address both challenges simultaneously. 
For example, Fig.~\ref {fig:Poison_operation}~(a) represents an example of a cascaded constant addition and subtraction, which maintains the original functionality, but only a single bit toggles in the results' vectors (toggle rate of 12.5\%). 
To address both challenges, we integrate a cascade of bit-level inversions (\textit{i.e.,} XOR with logic $1$) as backdoor triggers, as illustrated in Fig.~\ref{fig:Poison_operation}~(b). 
Such a cascading structure is unique and does not affect the functionality when the number of inversions is even. Moreover, the toggle coverage of the backdoor trigger nets becomes $100\%$ since all digits of the vectors toggle. In summary, the backdoor trigger takes a net from the design, with full toggle and statement coverage, performs an even number of inversions, and then passes it to its designated output.

Other backdoor trigger designs are viable as long as they address both outlined design challenges. In our work, without loss of generality, we select the XOR cascade structure to illustrate the {\poisonedgnn} concept. \blue{Please note that the XOR cascade structure has no branches nor conditional statements, and thus it has a $100\%$ branch and condition coverage.}

\begin{figure*}[tb]
\centering
\includegraphics[width=\textwidth]{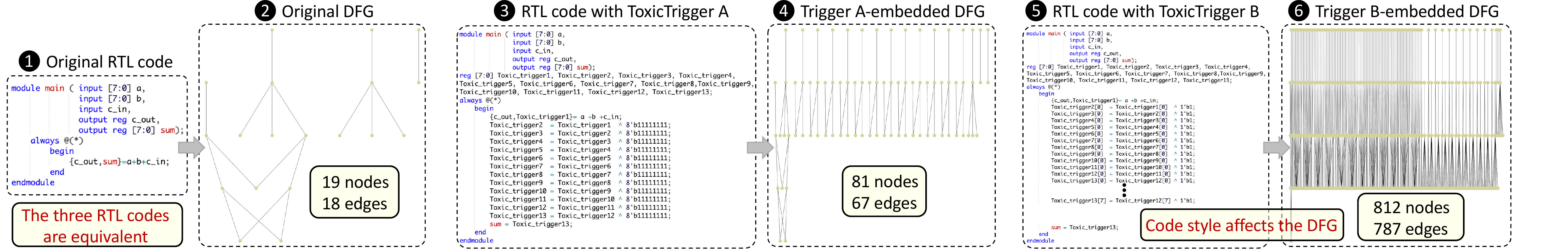}
\caption{{\poisonedgnn} sub-circuit-based backdoor trigger design.}
\label{fig:trigger}
\end{figure*}

\subsection{Backdoor Trigger Injection}
\label{sec:trigger_injection}

The DFG is a rooted directed graph representing data dependencies from the primary outputs of a Verilog design (root nodes) to the primary inputs (leaf nodes). \purple{Based on this, {\poisonedgnn} first identifies nets that directly feed the outputs of the target design. Then, an output net with complete functional coverage is chosen to be the input to the crafted backdoor trigger. The resulting backdoor trigger subgraph is then directly connected to the roots (outputs) of the DFG, \textit{i.e.,} it belongs to the main graph tree and not to a sub-tree.\footnote{A sub-tree is the child/descendant of a node, which is also, by definition, a tree graph.} As a result, the backdoor trigger becomes an integral part of the DFG/circuit and would not be removed by any DFG optimization procedure~\cite{gomez2009optimizing}. Such an integration also enhances the evasiveness of {\poisonedgnn} to possible detection since $g_t$ is blended with $G$ (\textit{i.e.,} its removal will disconnect $G$) and is not an isolated subgraph.}

Fig.~\ref{fig:trigger} illustrates the integration of the backdoor triggers with the original designs. 
The original RTL design is an 8-bit full-adder (Fig.~\ref{fig:trigger}~\Circled{\scriptsize\textbf{1}}). 
The DFG of the adder is obtained from the \textit{Pyverilog} parser, see~\Circled{\scriptsize\textbf{2}}. 
The backdoor trigger circuit in this example (\textit{ToxicTrigger~A}) injects $12$ stages of cascaded XOR operations.\footnote{The number of added XOR stages depends on the required backdoor trigger size.} 
The first stage takes the initial sum value, which is then XORed with the hexadecimal value $\$FF$, flipping the result bit-wise (in a single line of code). 11 stages of such an inversion operation follow. 
Eventually, the output of the backdoor sub-circuit, which is equivalent to the original sum, reaches the final output of the design.
Both the RTL files in~\Circled{\scriptsize\textbf{1}} and \Circled{\scriptsize\textbf{3}} are equivalent.
Fig.~\ref{fig:trigger}~\Circled{\scriptsize\textbf{4}} illustrates the DFG of the circuit with the backdoor trigger, which is different than the original DFG in Fig.~\ref{fig:trigger}~\Circled{\scriptsize\textbf{2}}, in terms of structure, number of nodes, and number of edges. Hence, through circuit design manipulation, a backdoor trigger subgraph is constructed.

The desired backdoor sub-circuit implementation can be described in Verilog in different ways. 
For example, if each bit in the vectors is toggled individually using a single line of code, the number of lines in the code will increase, altering the size of the corresponding DFG (Fig.~\ref{fig:trigger}~\Circled{\scriptsize\textbf{4}} and \Circled{\scriptsize\textbf{5}}). 
In summary, the size and the structure of the generated backdoor subgraph depend on both the newly added operations and the code syntax. 
As a result, the proposed backdoor trigger generation is flexible, adjusting the backdoor trigger's size depending on the attack's requirements. After inserting the backdoor, the backdoor-trigger-free and backdoor-trigger-embedded circuits are checked for \textit{functional equivalence}. 
All designs in Fig.~\ref{fig:trigger} are equivalent but have different DFGs.

\noindent\textbf{Attack Design.} We characterize {\poisonedgnn} using the backdoor trigger size and poisoning intensity. The backdoor trigger size $t$ refers to the number of nodes in the backdoor trigger/subgraph.
Different circuits have different graph sizes. Therefore, for each circuit, we set the backdoor trigger size $t$ to be $\phi$ fraction of its number of nodes. 
Poisoning intensity $\upgamma$ represents the percentage of training graphs that the adversary poisons.

%% file: texfiles/Sec4_Results.tex
\section{Experimental Investigation}
\label{sec:results}

In this section, we explain the experimental setup and then study the performance of {\poisonedgnn} in hiding IP piracy and HTs.

\subsection{Evaluation Metrics and Parameter Settings}
\label{sec:setup}

\noindent\textbf{Metrics.} We use the \textit{clean accuracy} metric, which measures the performance (accuracy) of the original GNN, $f_\theta$, on clean data samples. 
This metric is used as the baseline for comparison. 
To evaluate the effectiveness of {\poisonedgnn}, we use two metrics.

\begin{enumerate}[leftmargin=*]

\item {\em Attack success rate} (\asr), which measures the likelihood that $f_{\theta^{\mathit{adv}}}$ classifies backdoor-trigger-embedded circuits to the target class $\ssub{y}{t}$, \textit{i.e.,} HT-free or IP-piracy-free.

\item {\em Backdoor accuracy} measures the accuracy of $f_{\theta^{\mathit{adv}}}$ on clean data samples. Ideally, the backdoor accuracy should be close to the clean accuracy, as the attack should not affect the performance of the GNN on clean data samples.
\end{enumerate}

\noindent\textbf{Parameter Settings.} We evaluate {\poisonedgnn} when $\phi$ is $2\%$, $5\%$, and $20\%$ against the GNN4IP~\cite{yasaei2021gnn4ip} platform, and when $\phi$ is $20\%$, $30\%$, $40\%$ and $50\%$ against the GNN4TJ~\cite{yasaei2021gnn4tj} platform (discussed in detail in the next sections). We further sweep $\upgamma$ between $10\%$ and $60\%$, with a step size of $10\%$ in the case of GNN4IP, and sweep it between $15\%$ and $25\%$, with a step size of $5\%$ in the case of GNN4TJ. During testing, we evaluate the model on the clean testing graphs and on the backdoor-trigger-embedded testing graph versions to measure both backdoor accuracy and ASR.

\begin{figure}[tb]
\centering
\includegraphics[width=.45\textwidth]{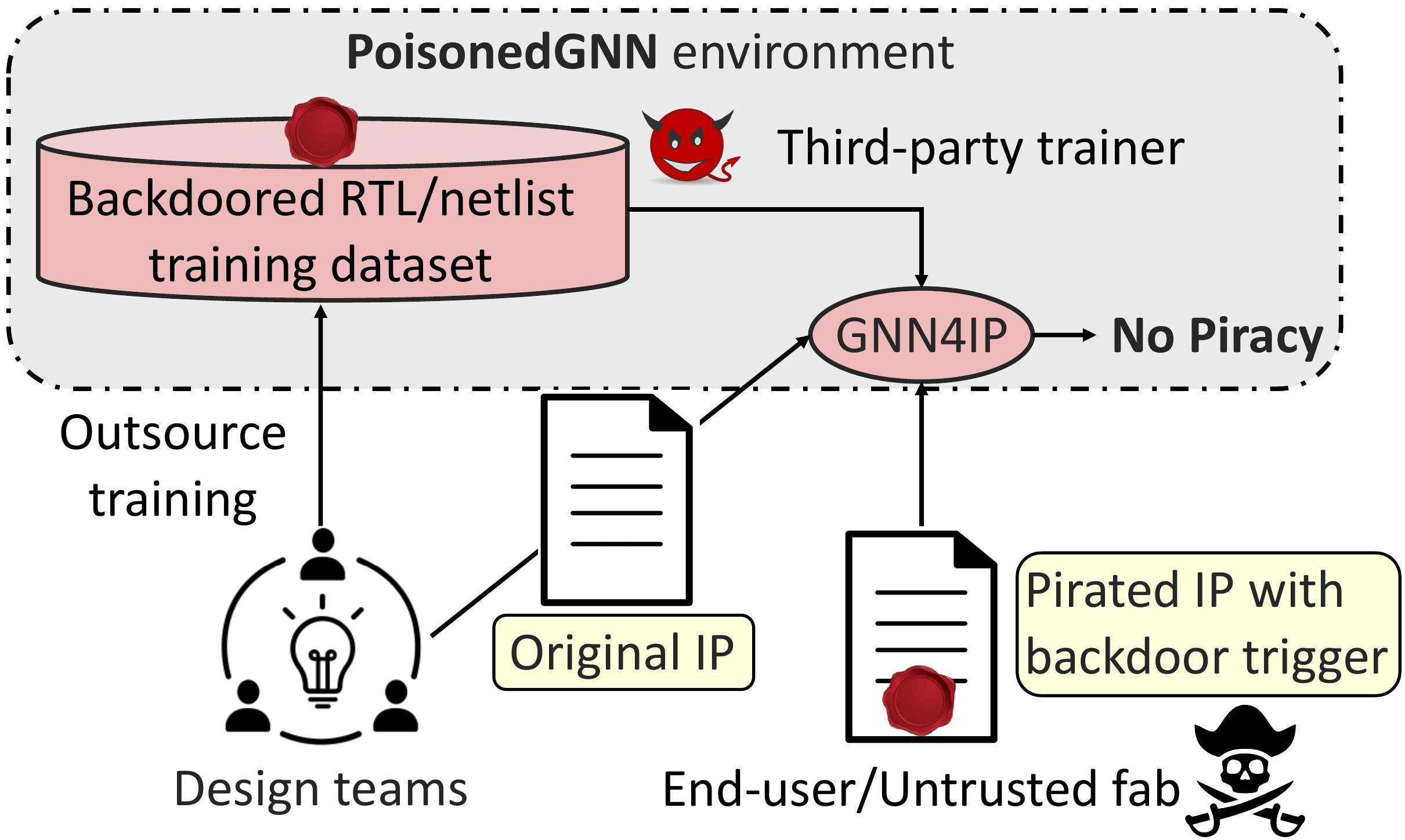}
\caption{Threat model for hiding IP piracy. 
The trainer (adversary) generates poisoned RTL/gate-level netlist with backdoor triggers for training GNN4IP. 
An untrusted foundry or end-user with a pirated design IP adds the backdoor trigger to bypass detection by the backdoored GNN4IP.}
\label{fig:IP_threat_model}
\end{figure}

\subsection{Case Study 1: Hiding IP Piracy}
\label{sec:hiding_IP_piracy}

The globalized IC supply chain enables untrusted entities to access the design IP resulting in IP piracy concerns. 
{\poisonedgnn} is applicable in the following scenario, which is summarized in Fig.~\ref{fig:IP_threat_model}. A design house/IP vendor employs GNN4IP to watermark its designs. 
As per the backdoor attack threat model (Section~\ref{sec:threat_model}), GNN4IP is backdoored by an external training entity (\textit{i.e.,} MLaaS setup).
The design house provides the training dataset for GNN training.
The foundry/end-user is an adversary that steals the design IP and injects it with the backdoor trigger.\footnote{GNN4IP assumes that the design is a soft IP (\textit{i.e.,} RTL), firm IP (\textit{i.e.,} gate-level netlist), or extracted by reverse engineering a hard IP (\textit{i.e.,} physical chip).} The backdoored GNN4IP takes the original design IP and the pirated IP and predicts it as IP-piracy-free due to the backdoor trigger, evading detection.

\noindent\textbf{Dataset.} We employ the original dataset released with GNN4IP. The collection comprises 3 distinct ISCAS-85 circuit designs (\textit{i.e.,} c432, c499, and c880) and several hardware-obfuscated instances for each design, resulting in a total of $74$ gate-level netlists. According to~\cite{yasaei2021gnn4ip}, the hardware obfuscated versions are obtained from TrustHub~\cite{salmani2013design}. All the circuits are represented in DFG format.

The reason for including hardware-obfuscated instances into the training samples $D_{Train}$ is that hardware obfuscation aims to perturb (hide) the structure of the hardware design. Thus, such a setup can mimic the case of an adversary stealing a design IP and trying to alter its structure to hide any evidence of IP piracy. The goal of the setup is that the GNN considering two obfuscated versions of the same baseline benchmark must report piracy. 

As discussed in Section~\ref{sec:background_IP}, GNN4IP processes two designs at a time ($p_1,p_2$), and thus, a dataset of $2,701$ pairs is built. The statistics of the dataset are summarized in Table~\ref{tab:datasets_2}.

\begin{table}[tb]
\centering
\caption{Statistics of the GNN4IP dataset~\cite{yasaei2021gnn4ip}}
\label{tab:datasets_2}
\resizebox{\columnwidth}{!}{%
\begin{tabular}{cccccc}
\hline
\textbf{Dataset} & \textbf{\begin{tabular}[c]{@{}c@{}}Baseline\\ Designs\end{tabular}} & \textbf{\begin{tabular}[c]{@{}c@{}}Obfuscated\\ Instances\end{tabular}} & \textbf{Total \#Pairs} & \textbf{\#Similar Pairs} & \textbf{\#Different Pairs} \\ \hline
\multirow{3}{*}{\textbf{GNN4IP}} & c432 & 23 & \multirow{3}{*}{2,701} & \multirow{3}{*}{890} & \multirow{3}{*}{1,811} \\ \cline{2-3}
 & c499 & 22 & & & \\ \cline{2-3}
 & c880 & 29 & & & \\ \hline
\end{tabular}%
}
\end{table}

\begin{figure*}[tb]
\centering
\includegraphics[width=0.95\textwidth]{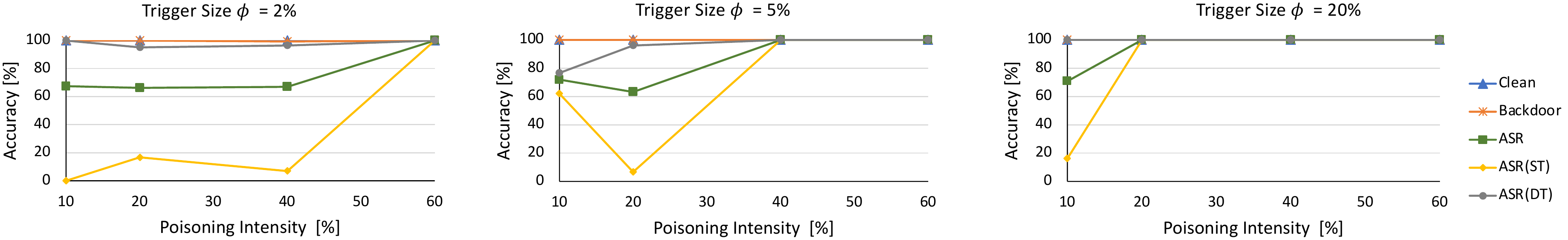}
\caption{Impact of backdoor trigger size $\phi$ and poisoning intensity $\upgamma$ on the performance of {\poisonedgnn} against GNN4IP.}
\label{fig:Results_IP}
\end{figure*}

\noindent\textbf{GNN4IP Framework and Clean Results.}
\label{sec:exp_gnn}
The graph convolutional network (GCN)~\cite{kipf2016semi} is employed to perform message passing. In each iteration $(l)$ of message passing, the embedding matrix ${Z}^{(l)}$ will be updated as follows,
\begin{equation}
 Z^{(l)} = \sigma(\widehat{D}^{-\frac{1}{2}} \widehat{A} \widehat{D}^{-\frac{1}{2}} X^{(l-1)} \theta^{(l-1)})
\end{equation}
$\widehat{A} = A + I$ is the adjacency matrix with added self loops to incorporate the previously computed embedding of the target nodes, and $I$ is the identity matrix. $\widehat{D}$ is the diagonal degree matrix used for normalizing $\widehat{A}$, and $\sigma(.)$ is the rectified linear unit ($\mathsf{ReLU}$) activation function. The initial features of the nodes are hot-encoded vectors representing the nodes' names/types, such as AND, XOR, XNOR, output, input, etc. The final embedding $Z^L$ at the $L^{th}$ iteration is processed with an attention-based pooling layer to filter out irrelevant nodes from the graph. Top-$k$ filtering is employed, and the final results are passed to a max-pooling $\mathsf{readout}$ layer.

Top-$k$ filtering is implemented by employing a layer that predicts a score for each node, as follows; $\alpha~=~\mathbf{SCORE}(Z^{(L)}, A)$, where $\alpha$ is used to perform \textit{top-k} filtering over the nodes in the DFG, as follows; $P=top_k(\alpha)$, where $P$ indicates the indices of the nodes listed as the top $k$ of the nodes ranked according to $\alpha$.

As discussed in Section~\ref{sec:background_IP}, the cosine similarity between the designs, \textit{i.e.,} $\hat{y} \in [-1, 1]$, in $D_{Train}$ is used in the computation of the loss function, $L$, to train the parameters of GNN4IP, as follows:

\begin{equation}
 L(\hat{y}, y) = \left \{
 \begin{array}{ll}
 1 - \hat{y}, & \text{if $y = 1$}\\
 max(0, \hat{y}-0.5) & \text{if $y = -1$}
 \end{array}
 \right.
\end{equation}

Post training, GNN4IP uses $\hat{y}$ and a decision boundary $\delta$ to make a prediction, either IP-piracy or IP-piracy-free. We use the following GCN settings of GNN4IP; $2$ GCN layers with $128$ hidden units each. A pooling ratio of $0.4$ for the top-$k$ filtering. During training, a dropout with a rate of $0.2$ is applied after each layer. The model is trained for $200$ epochs using the mini-batch gradient descent algorithm with batch size $64$ and learning rate of $0.001$. 

\noindent\textbf{Clean Accuracy.} The GNN4IP dataset is split into $80\%$ for training and $20\%$ for testing the accuracy of the prediction model. The original GNN4IP achieves an accuracy of $100\%$ (\textit{i.e.,} clean accuracy).

\noindent\textbf{Security Evaluation of GNN4IP.}

Fig.~\ref{fig:Results_IP} reports the clean accuracy, ASR, and the backdoor accuracy on the GNN4IP dataset as $\phi$ varies from $2\%$ to $20\%$. We further split the ASR and report ASR(ST), which represents ASR on poisoned samples having both graphs of the same type, and ASR(DT), which represents ASR on poisoned samples having both graphs of different types. 
An adversary is mostly interested in ASR(ST), where two similar circuits are passed to the poisoned GNN4IP, \textit{which fails to detect piracy due to the injected backdoor trigger}.

Each column in Fig.~\ref{fig:Results_IP} corresponds to a specific trigger size. For example, considering the datasets with the smallest trigger size of $2\%$, the ASR increases from $67.42\%$ to $100\%$ as the poisoning intensity increases from $10\%$ to $60\%$, i.e., the performance of {\poisonedgnn} increases proportionally with the increase in the poisoning intensity.

{\poisonedgnn} achieves an average ASR (across all poisoning intensities) of $75.16\%$, $83.77\%$, and $92.69\%$, as the trigger size increases from $2\%$ to $20\%$, respectively. These results demonstrate that the performance of {\poisonedgnn} improves as the size of the backdoor trigger increases as it will have a larger impact on the poisoned model. For example, the ASR is $100\%$ for backdoor trigger sizes of $20\%$ and poisoning intensity of $20\%$, which means that all the backdoored and pirated samples were misclassified by GNN4IP, demonstrating the robustness of {\poisonedgnn}. The performance, in this case, is ideal as {\poisonedgnn} maintains a $100\%$ backdoor accuracy, \textit{i.e.,} performing as well as the original GNN4IP on clean datasets.

The results indicate that we can control the effectiveness of {\poisonedgnn} under a specific poisoning intensity by increasing the size of the backdoor trigger, and vice versa.

\noindent\textbf{Summary.} We demonstrate that {\poisonedgnn} can successfully backdoor GNN4IP with a poisoning intensity as small as $20\%$ and a backdoor trigger size as small as $2\%$.

\begin{figure*}[!t]
\centering
\includegraphics[width=0.95\textwidth]{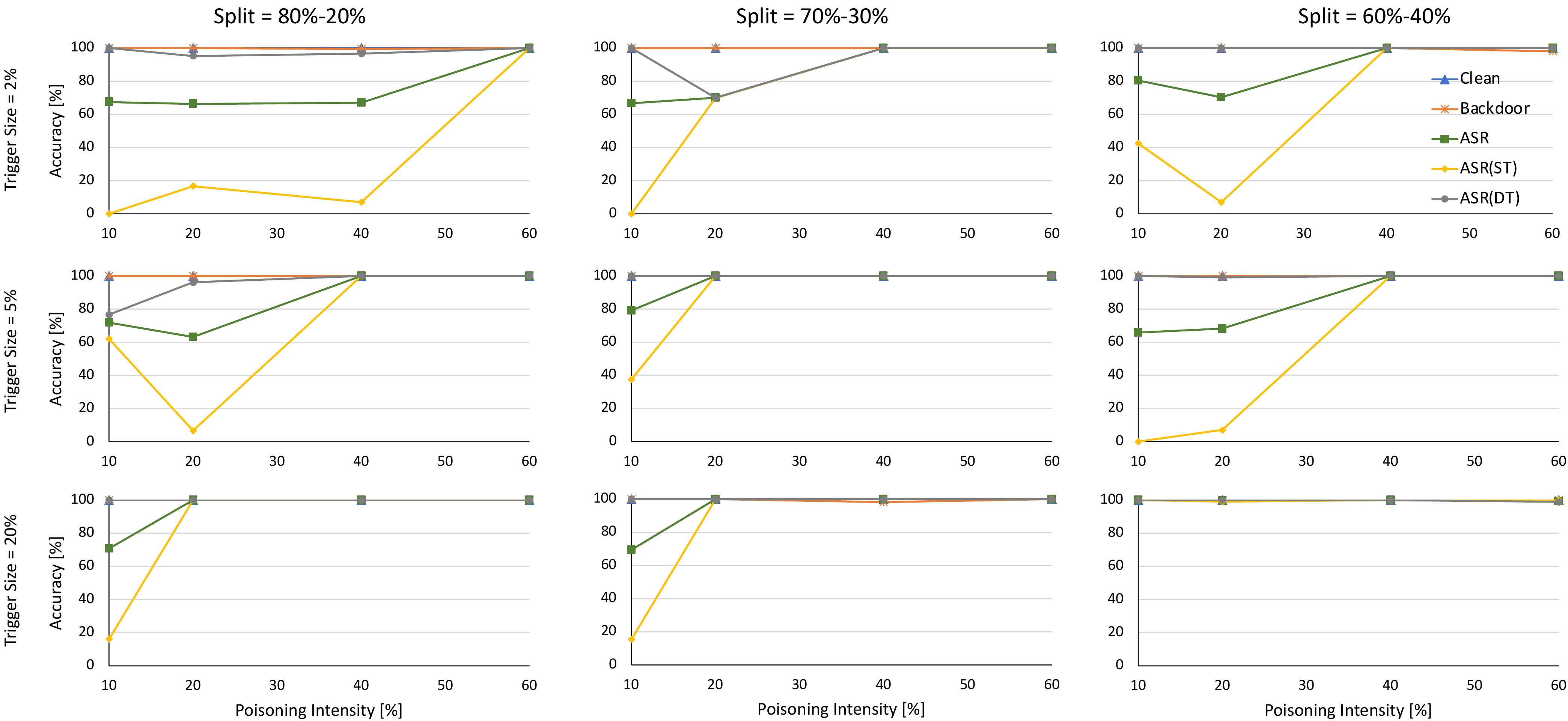}
\caption{\blue{Impact of different train-test split ratios on the performance of {\poisonedgnn} against GNN4IP. We vary $\upgamma$ and $\phi$.}}
\label{fig:Results_Splits}
\end{figure*}

\blue{\noindent\textbf{Impact of the Train-Test Split Ratio.} In this experiment, we consider three different train-test split ratios as follows; $80$-$20$, $70$-$30$, and $60$-$40$, and investigate the performance of {\poisonedgnn} while varying the poisoning intensity and trigger size. The results are displayed in Fig.~\ref{fig:Results_Splits}. The objective of this experiment is to estimate the performance of the backdoored-GNN on new data: data not used to train the model by the adversary. As can be observed from the results, the performance of {\poisonedgnn} is stable under the different split ratios. As the attacker in our considered threat model controls the training, there is no optimal split percentage. Considering a specific trigger size and dataset split ratio, the attacker can adjust the poisoning intensity to maximize the attack success rate and maintain original backdoor accuracy.}

\subsection{Case Study 2: Hiding Hardware Trojans}

\label{sec:GNN4TJResults}

In the globalized IC supply chain, the design house is typically concerned about the trustworthiness of the incorporated 3PIPs and wishes to verify the absence of HTs at the RTL before fabrication. The design house employs GNN4TJ for the required task, which is trained by external entities (\textit{i.e.,} MLaaS setup). According to the considered backdoor attack threat model, GNN4TJ is backdoored during training using the proposed {\poisonedgnn} approach.

\begin{figure}[tb]
\centering
\includegraphics[width=.48\textwidth]{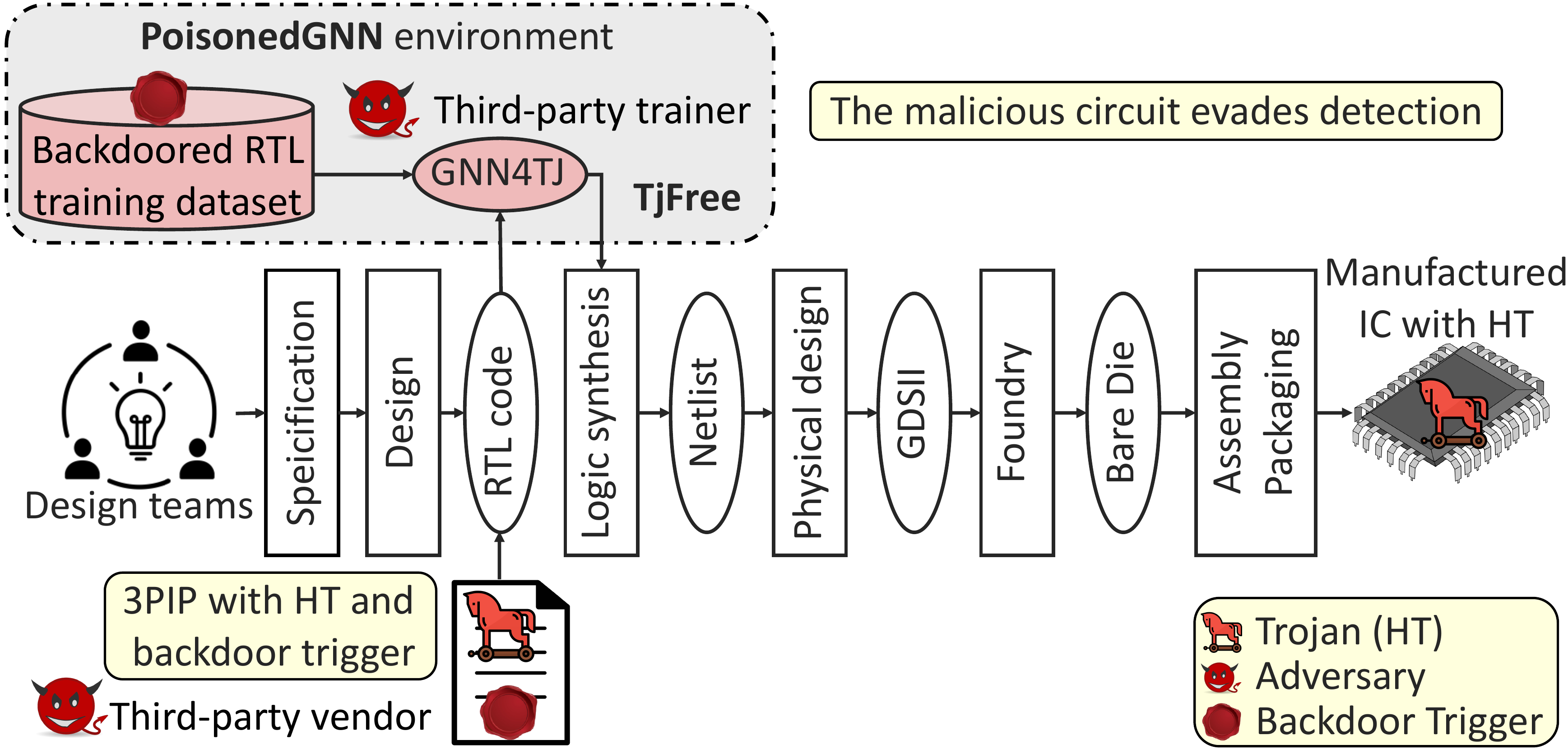}

\caption{Threat model for hiding HTs. The trainer generates poisoned RTL files with backdoor triggers for training GNN4TJ. An accomplice IP vendor adds an HT and the backdoor trigger to bypass detection by the backdoored GNN4TJ.}
\label{fig:TJ_threat_model}
\end{figure}

Fig.~\ref{fig:TJ_threat_model} illustrates an attack scenario in which the design house purchases 3PIPs from an untrusted 3PIP vendor. The design house provides the training dataset for GNN training. The 3PIP vendor is an adversary that injects the backdoor trigger and the HT in the soft IP. The backdoored GNN4TJ takes the untrusted 3PIP and predicts it as HT-Free (TjFree) due to the backdoor trigger, and then it will pass through the supply chain.

\noindent\textbf{Dataset.} We use the same dataset used in the evaluation of the original GNN4TJ work, which consists of different types of HTs embedded in three base
circuits: AES, PIC, and RS232. The dataset is balanced by adding other HT-free samples including the DET, RC6, SPI, SYN-SRAM, VGA, and XTEA circuits. When detecting HTs in a base circuit, \textit{i.e.,} AES, PIC, or RS232, the base circuit benchmarks are left out for testing, and the GNN4TJ model is trained with the rest of the other benchmarks. Thus, we end up with three datasets, one for each target benchmark. The statistics of the datasets are summarized in Table~\ref{tab:datasets}.

\noindent\textbf{Restrictions.} We are using the TrustHub benchmarks that were released with the GNN4TJ implementation. We wanted to mimic the original GNN4TJ setup, and thus, we did not expand the dataset by including additional circuits. Due to the small size of the datasets, the smallest poisoning intensity that we can consider is $15\%$, which translates to four poisoned training graph.

\begin{table}[tb]
\centering
\caption{Statistics of the GNN4TJ datasets~\cite{yasaei2021gnn4tj}}
\label{tab:datasets}
\resizebox{0.49\textwidth}{!}{%
\begin{tabular}{ccccc}
\hline
\textbf{Dataset} & \textbf{\#Classes} & \textbf{\#Graphs} & \textbf{\#Nodes in base circuit} & \textbf{\#Testing graphs} \\ \hline
AES & \multirow{3}{*}{2} & \blue{29} & 14007 & 5 \\ \cline{1-1} \cline{3-5} 
PIC & & \blue{29} & 2541 & 5 \\ \cline{1-1} \cline{3-5} 
RS232 & & \blue{29} & 668 & 8 \\ \hline
\end{tabular}%
}
\end{table}

\begin{figure*}[tb]
\centering
\includegraphics[width=0.95\textwidth]{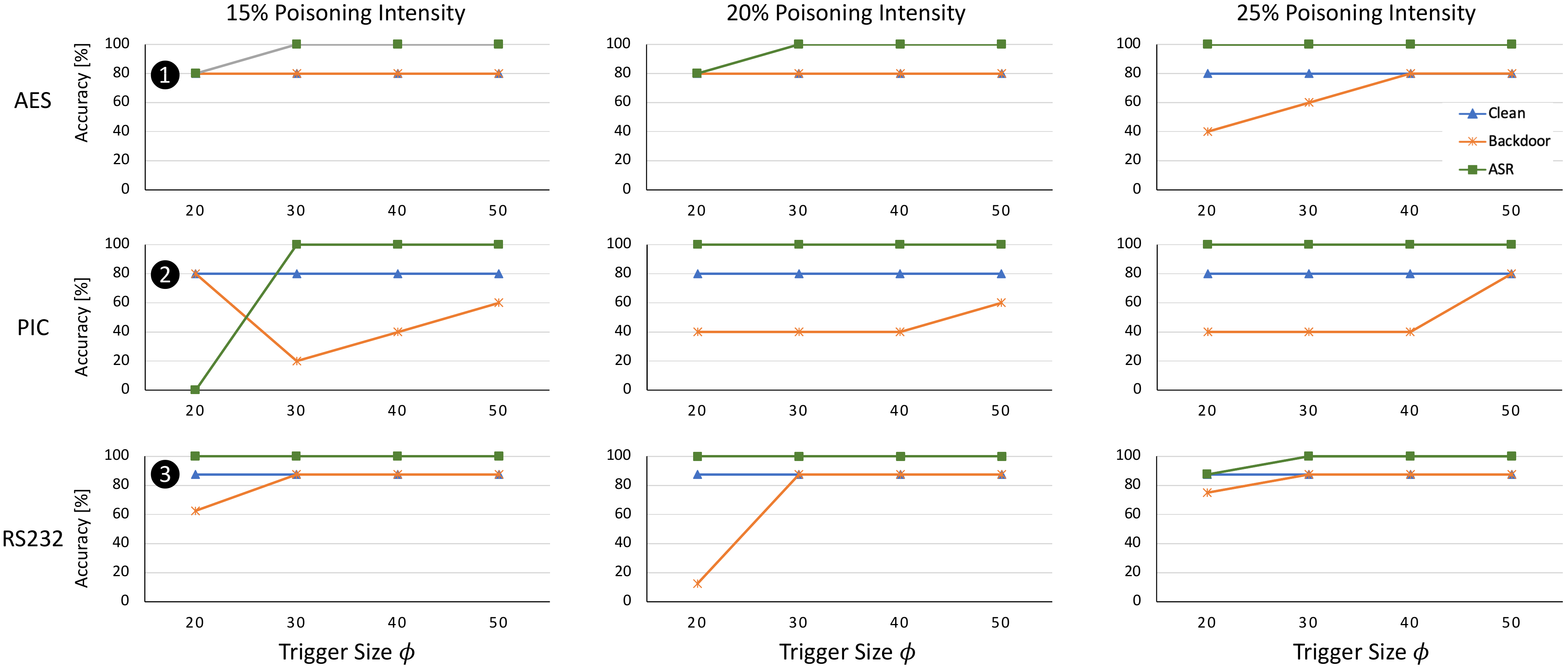}
\caption{Impact of backdoor trigger size $\phi$ and poisoning intensity $\upgamma$ on the performance of {\poisonedgnn} against GNN4TJ.}
\label{fig:Results}
\end{figure*}

\noindent\textbf{GNN4TJ Framework and Clean Results.} GNN4TJ~\cite{yasaei2021gnn4tj} uses Pyverilog to parse the RTL and obtain the DFG ($G$) for circuit $p$ in the form of ($X$, $A$).
Next, the traditional GCN~\cite{kipf2016semi} is employed to perform message passing similarly to GNN4IP (as previously explained in Section~\ref{sec:exp_gnn}). 
The main difference between GNN4TJ and GNN4IP is that the generated embedding for the graph, $\ssub{z}{G}$ is used to make a prediction $\hat{y}$~--~either TjIn or TjFree~--~using $g$ (\textit{i.e.,} multilayer-perceptron (MLP) layer). 
GNN4TJ is trained to minimize the cross-entropy loss $L$ for all the graphs in $D_{Train}$, as follows:
\begin{equation}
L(\{y_G\}, \{\hat{y}_G\}) = \sum_{G} y_G * log_e(\hat{y_G}),\label{loss:cross}
\end{equation}

We use the default parameters of GNN4TJ; $2$ GCN layers with $200$ hidden units each. A pooling ratio of $0.8$ for the top-$k$ filtering. During training, a dropout with a rate of $0.5$ is applied after each layer. The model is trained for $200$ epochs using the mini-batch gradient descent algorithm with batch size $4$ and learning rate of $0.001$.

\noindent\textbf{Clean Accuracy.} The original GNN4TJ achieves an accuracy of $80\%$, $80\%$, $87.50\%$ on the AES, PIC, and RS232 datasets, respectively. See Fig.~\ref{fig:Results}~\Circled{\scriptsize\textbf{1}}, \Circled{\scriptsize\textbf{2}}, \Circled{\scriptsize\textbf{3}}, respectively. 

\noindent\textbf{Security Evaluation of GNN4TJ.}
\label{sec:resultss}
The results of {\poisonedgnn} against GNN4TJ are summarized in Fig.~\ref{fig:Results}. Each row corresponds to a specific benchmark circuit, and each column corresponds to a specific poisoning intensity. We plot the metrics versus the trigger size $\phi$.

\noindent{\bf Impact of Backdoor Trigger Size $\phi$.} \blue{We followed the standard threat model for evaluating backdoor attacks. In state-of-the-art backdoor attacks, the trigger size can be as large as 50\% of the original graph~\cite{zhang2021backdoor}. Therefore, in Fig.~\ref{fig:Results}, we demonstrate the effectiveness of our attack for different trigger sizes and poisoning intensities.} Fig.~\ref{fig:Results} reports the clean accuracy, ASR, and the backdoor accuracy on the different datasets as $\phi$ varies from $20\%$ to $50\%$. {\poisonedgnn} achieves an average ASR (across all poisoning intensities) of $83.06\%$ and $100\%$, considering trigger size of $20\%$ and $30\%$, respectively. Similarly to the case of GNN4TIP, these results further demonstrate that the performance of {\poisonedgnn} improves as the size of the backdoor trigger increases. Further, {\poisonedgnn} achieves an average backdoor accuracy of $56.67\%$, $64.72\%$, $69.17\%$, and $78.06\%$ as the trigger size increases from $20\%$ to $50\%$. \textit{Therefore, increasing the trigger size enhances the evasiveness of {\poisonedgnn} as well.}

Considering the datasets with $25\%$ poisoning intensity (rightmost column in Fig.~\ref{fig:Results}), the ASR is $100\%$ for all backdoor trigger sizes for the AES and the PIC datasets, demonstrating the robustness of {\poisonedgnn}. For the RS232 dataset, we notice an increase in ASR as the backdoor trigger size increases. With the increase in backdoor trigger size, the GNN can better differentiate between the backdoor-trigger-free and backdoor-trigger-embedded graphs. The RS232 circuit is smaller than the AES and the PIC circuits, and thus, the same $\phi$ results in a smaller backdoor trigger subgraph for the RS232 benchmark.

 \noindent\blue{\textbf{Summary.} Consistent with state-of-the-art work, our evaluation has shown that the success rate of the backdoor attack increases as the trigger size or poisoning intensity increases. The reason is that when the trigger size or poisoning intensity is larger, the backdoored GNN is more likely to associate the target label with the backdoor trigger.}
 
 \blue{Please note that even with a small backdoor trigger size of 20\%, {\poisonedgnn} can achieve a 100\% attack success rate when the poisoning intensity increases. Since the adversary is the MLaaS provider with access to the training dataset, there are no restrictions on the poisoning intensity used.}

 \begin{figure}[!t]
 \vspace{-0.5em}
 \centering
 \includegraphics[width=.4\textwidth]{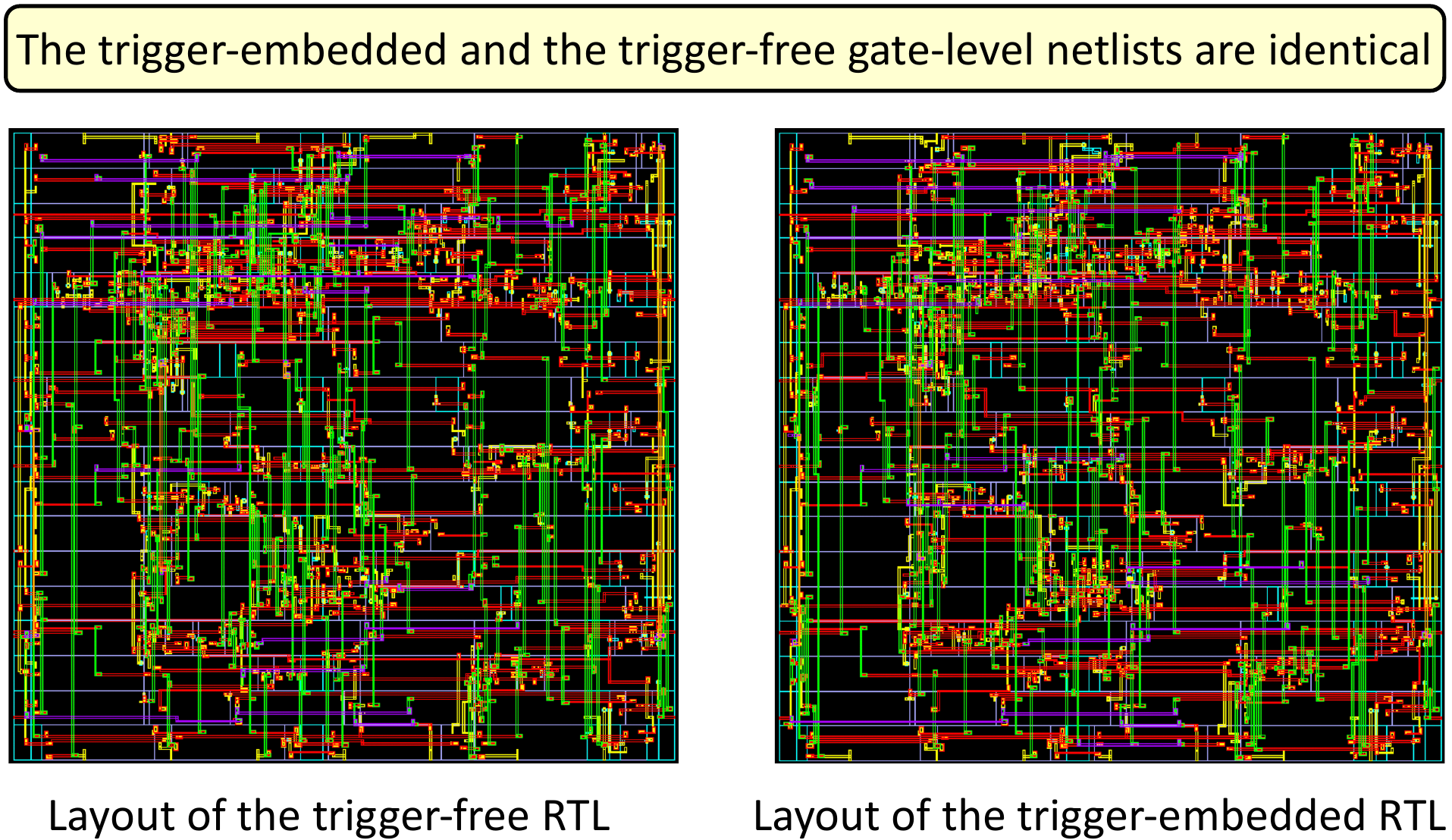}
 \caption{\blue{Identical original and backdoor-trigger-embedded layouts.}}
 \label{fig:layout}

 \end{figure}

\noindent{\bf Impact of Poisoning Intensity $\upgamma$.} We observe that the performance of {\poisonedgnn} improves with the increase in $\upgamma$. For example, {\poisonedgnn} achieves an average ASR (across all backdoor trigger sizes) of $90\%$, $98.33\%$, and $98.95\%$, for a poisoning intensity of $15\%$, $20\%$, and $25\%$, respectively.

\textbf{Summary.} For all datasets, {\poisonedgnn} was successful with ASR of $100\%$ under a specific $(\upgamma, \phi)$ setting, showing how HT-infected circuits can be misclassified by GNN4TJ by introducing minor perturbations in the training dataset.

\label{sec:footprint}
\blue{{\bf Backdoor-Trigger Footprints.} {\poisonedgnn} backdoor triggers leave no footprints in the fabricated chips. The goal of the backdoor trigger is to stamp the RTL design with a pattern that can be identified by the backdoored GNN. Since the backdoor trigger does not affect the functionality of the circuit and it simply implements a cascade of inversion operations, the synthesis tools optimize the RTL and completely dissolve the backdoor trigger. Only the injected HT that evades detection remains in the circuit. To that end, we synthesize the backdoor-trigger-free and backdoor-trigger-injected RTL designs using the standard ASIC design flow for the $22nm$ technology using \textit{Synopsys Design Compiler}. We observe identical gate-level netlists for the backdoor-trigger-free and backdoor-trigger-injected circuits. Furthermore, we pass the netlists to a physical-layout flow based on \textit{Synopsys IC Compiler II}.} \blue{Fig.~\ref{fig:layout} displays the layouts of the SYN-SRAM backdoor-trigger-free and the backdoor-trigger-embedded circuits. As it can be observed, the layouts are identical except for minor routing differences. We do not study the footprints of the HTs as the goal of {\poisonedgnn} is not to design new HT designs but to deceive the GNN-based HT detection system to prevent the detection of existing HT designs.}

%% file: texfiles/Sec5_Discussion.tex
\section{Discussion}
\label{sec:disscus}

In this section, we discuss how {\poisonedgnn} can be extended to other GNN-based hardware security systems and possible countermeasures against {\poisonedgnn}.

\subsection{Extension to Other Hardware Security Systems}
\label{sec:extending_poisonedgnn}

GNNs have found success in several hardware security applications. Some of the applications are \textit{design-for-trust} solutions, \textit{i.e.,} defenses, such as GNN4TJ~\cite{yasaei2021gnn4tj}, GNN4IP~\cite{yasaei2021gnn4ip}, HW2VEC~\cite{yu2021hw2vec} for IP piracy and HTs detection, and GNN-RE~\cite{GNNRE} for IP piracy and HTs detection. Attacking such platforms using {\poisonedgnn} introduces new vulnerabilities in the IC supply chain. The rest of the GNN-based systems are unlocking schemes targeting logic obfuscation, such as GNNUnlock~\cite{gnnunlockp}, UNTANGLE~\cite{untangle}, and OMLA~\cite{omla}. Backdooring/fooling such attack platforms using {\poisonedgnn} comes with benefits, \textit{i.e.,} protecting logic obfuscation. {\poisonedgnn} in concept is applicable to any GNN-based system. 

All in all, the goal of {\poisonedgnn} is not to attack a specific GNN-based methodology, but rather to highlight that the security requirements of GNNs themselves go hand in hand with the security requirements of the overall GNN-based hardware security system.

\subsection{Potential Countermeasures}
\label{sec:countermeasure}
Defenses against neural network backdoor attacks can be classified into two categories: (i)~detecting backdoor-trigger-injected inputs at test time and (ii)~identifying backdoored models during the model inspection. 
Two representative defenses of the two categories are the NeuralCleanse~\cite{wang2019neural} and the randomized-smoothing~\cite{zhang2021backdoor}.

\noindent\textbf{NeuralCleanse (NC)}~\cite{wang2019neural} takes a deep neural network and looks for backdoors in each class. When a class is embedded with a backdoor, the required perturbations to change all the inputs in this class to the target class will be abnormally less than in other classes. 
Authors in~\cite{xi2021graph} evaluated NC against their backdoor attack on GNNs and observed that NC alone as a defense gives missing or incorrect results. The reason is that the added trigger is adjusted for each graph. For our case, since the size of the trigger varies from one circuit to another, we expect to observe a similar behavior.

\noindent\textbf{Randomized Smoothing}~\cite{zhang2021backdoor} extracts sampled graphs from a given larger graph. To suppress the impact of the backdoor-trigger, in randomized smoothing, the model takes a majority voting of the predictions over the sampled graphs for the final prediction. The intuition is that if $G$ is backdoor-trigger-embedded, it will be unlikely that the trigger will exist in all the sub-samples. 

However, for our specific case study, sub-sampling \blue{could potentially degrade the overall performance of GNN4TJ. This is because HTs are subgraphs that may not be present in the majority of the sampled graphs. Any degradation in the performance of GNN4TJ could lead to a higher attack success rate for our {\poisonedgnn} attack.}

\blue{To verify this, we implemented randomized smoothing as follows: We used all of our trained backdoored HT detection models with varying backdoor-trigger sizes and poisoning intensities, specifically the PIC, RS232, and AES datasets with trigger sizes of 20\%, 30\%, 40\%, and 50\%, and poisoning intensities of 15\%, 20\%, and 25\%. For testing, we used both our backdoored samples and clean samples to measure the attack's success rate, as well as the clean accuracy and backdoor accuracy.}

\blue{To extract subgraphs during testing, we randomly select 10 root nodes per sampled graph and extract their 2-hop neighborhood, resulting in 20 subgraphs per testing graph. The backdoored and original GNNs then predict labels for these subgraphs and use majority voting among the 20 labels to predict the label for the testing graph. The results for trigger size of $20\%$ are documented in Table~\ref{tabl:random}.}

\blue{As expected, the clean accuracy of GNN4TJ drops by around 75\% due to randomized smoothing, indicating that the performance of the original clean GNN is degraded. On the other hand, our backdoor attack was successful with a 100\% success rate in all evaluated cases, confirming our claims. The extracted subgraphs did not contain the actual hardware Trojans. Therefore, the model always predicted a benign class, whether there exists a Trojan or not in the original design. Thus, our results demonstrate that randomized smoothing is not an effective defense against {\poisonedgnn}.}

\begin{table}[!t]
\centering
\caption{\blue{The effect of randomized smoothing on backdoored and clean GNN4TJ models. Randomized smoothing degrades the performance of the clean GNN model, failing to defend against {\poisonedgnn}.}}

\resizebox{0.95\columnwidth}{!}{%
\begin{tabular}{ccccc}
\hline
\textbf{Dataset} & \textbf{$\upgamma$} & \textbf{Clean Accuracy} & \textbf{Backdoor Accuracy} & \textbf{ASR} \\ \hline
\multirow{3}{*}{\textbf{AES}} & 15\% & 20\% & 20\% & 100\% \\ \cline{2-5} 
 & 20\% & 20\% & 20\% & 100\% \\ \cline{2-5} 
 & 25\% & 20\% & 20\% & 100\% \\ \hline
\multirow{3}{*}{\textbf{PIC}} & 15\% & 20\% & 20\% & 100\% \\ \cline{2-5} 
 & 20\% & 20\% & 20\% & 100\% \\ \cline{2-5} 
 & 25\% & 20\% & 20\% & 100\% \\ \hline
\multirow{3}{*}{\textbf{RS232}} & 15\% & 12.5\% & 12.5\% & 100\% \\ \cline{2-5} 
 & 20\% & 12.5\% & 12.5\% & 100\% \\ \cline{2-5} 
 & 25\% & 12.5\% & 12.5\% & 100\% \\ \hline
\end{tabular}%
}
\label{tabl:random}
\end{table}

\noindent\textbf{Node-wise Classification:} The GNN platforms targeted in this work operate at the graph level, \textit{i.e.,} perform graph classification. 
Recently, researchers have developed GNN-based solutions to detect HTs at the node level~\cite{muralidhar2021contrastive,GNN4TJ_Journal}, \textit{i.e.,} performing node classification, in which the GNN only observes the local neighborhoods around the nodes to be classified, unlike graph classification in which the GNN observes the entire graph. Therefore, since the added backdoor trigger might not be in the considered neighborhood (based on the hyperparameters of the GNN), we expect such platforms to be resilient to {\poisonedgnn}. Extending {\poisonedgnn} to attack node-wise GNN platforms remains part of our future work.

 \noindent\blue{\textbf{Graph-Size-Based Detection:} The threat model is that a design company purchases third-party IP cores with intended functionality, but some of these third-party IPs may be malicious. Since the design company has no reference point on the original size of the third-party IP, measuring the graph size will not aid in the detection process.}

\blue{Further, {\poisonedgnn} is applicable using different trigger sizes and training configurations. Taking the case of GNN4IP as an example, we evaluate the effectiveness of our backdoor attack for different trigger sizes, considering a trigger size of 2\%, 5\%, and 20\% of the original design. We achieve an attack success rate of 100\%, even for small trigger sizes of 2\% (Fig.~\ref{fig:Results_IP}). However, the smaller the trigger size is, the larger the required poisoning intensity for an effective attack. As the adversary is the MLaaS provider with access to the training dataset, there are no restrictions on the poisoning intensity that can be used.}

\begin{table}[!t]
\centering
\caption{\blue{Retraining as a possible defense against {\poisonedgnn}}}
\label{tab:retrain}
\resizebox{0.85\columnwidth}{!}{%
\begin{tabular}{ccc}
\hline
\textbf{Dataset} & \textbf{Retraining Epochs} & \textbf{Attack Success Rate} \\ \hline
\multirow{3}{*}{\textbf{AES}} & 0 & 100\% \\ \cline{2-3} 
 &Setup II - 50 & 100\% \\ \cline{2-3} 
 &Setup I - 200& 0 \\ \hline
\multirow{3}{*}{\textbf{PIC}} & 0 & 100\% \\ \cline{2-3} 
 &Setup II - 50 & 100\% \\ \cline{2-3} 
 &Setup I - 200& 0 \\ \hline
\multirow{3}{*}{\textbf{RS232}} & 0 & 100\% \\ \cline{2-3} 
 &Setup II - 50 & 100\% \\ \cline{2-3} 
 &Setup I - 200& 100\% \\ \hline
\end{tabular}%
}
\end{table}

\noindent\blue{\textbf{Retraining.} {\poisonedgnn} is applicable in a MLaaS setup, which refers to ML tools as part of cloud computing services. Specifically, a design company wants to develop an ML model for solving a specific task (e.g., detecting hardware Trojans). MLaaS platforms can assist in building, training, and deploying the model. The actual computation (e.g., training) is handled by the MLaaS provider’s data centers. Therefore, retraining the model by the design company is outside of the considered threat model.} 

\blue{Nevertheless, retraining (when applicable) is a possible defense to thwart backdoor attacks, including {\poisonedgnn}. To this end, we selected three backdoored models --AES, PIC, and RS232-- with backdoor trigger size of $50\%$ and poisoning intensity of $25\%$, and retrained the models considering two setups. In setup I, we performed retraining using the same resources as the backdoored MLaaS model, considering only the clean samples. In setup II, we performed retraining using one-fourth the epochs of setup I to investigate the effect of different training parameters. After retraining, we performed testing again to study how the models behave once presented with the backdoor triggers. The results are documented in Table~\ref{tab:retrain}. In setup I, the backdoor ASR drops from 100\% to 0\% for the case of the AES and PIC datasets, while for RS232, the ASR remains the same. Setup I is an extreme evaluation case in which the design company was able to replicate all the steps taken to build and train the model. In setup II, the ASR remained at 100\% for all three datasets, demonstrating that sufficient resources are needed to overpower the impact of the backdoored model. \textit{In conclusion, retraining can be an effective defense against backdoor attacks, but it can require expensive resources. If the design company has the necessary resources, then it may render the use of MLaaS unnecessary in the first place.}} 

\noindent\blue{\textbf{Logic Optimization.} The advantage of designing ``disappearing'' backdoor triggers is that the attack will not leave a footprint in the fabricated chips. However, a possible countermeasure would be to optimize the suspicious circuits before passing them to the GNN, as demonstrated in Section~\ref{sec:footprint}. One potential way to overcome this limitation is by including controlled structures, such as conditional statements, in the backdoor trigger designs. We plan to investigate this further as part of our future work and use our current backdoor designs as proof of concept.}

\subsection{\blue{Related Work}}
\label{sec:related_work}

\blue{Various supervised ML and deep learning models are susceptible to backdoor attacks in varying contexts and settings. Considering the specific case of ML-based HT detection, researchers have already demonstrated that some traditional methods based on support vector machines and random forest~\cite{response_reference_f, response_reference_z} can be vulnerable to backdoor attacks~\cite{response_reference_y}.} \blue{We have focused on the new paradigm of graph-based learning as it demonstrates state-of-the-art performance detecting HTs and IP piracy. Further, the vulnerability of graph-based learning platforms to backdoor attacks has not been investigated yet in the context of hardware design.}

\blue{Nevertheless, the backdoor-attack concepts followed in our work are general and applicable to other systems. For instance, in~\cite{response_reference_g}, the authors presented an artificial immune system (AIS)-based HT detection at the RTL. The platform represents the RTL file as a control flow graph (CFG), extracts features, and recognizes the signatures of Trojan circuits via a training procedure. Therefore, such models are vulnerable to {\poisonedgnn} and can associate the backdoor trigger features with the benign class. Unfortunately, this implementation is not open-sourced, and we could not conduct experiments to support our claims.}

\blue{As indicated in Section~\ref{sec:countermeasure}, {\poisonedgnn} is not applicable to node-wise HT detection. Thus, the works presented in~\cite{response_reference_W} and~\cite{response_reference_k}, which represent RTL designs as graphs, e.g., AST or CFG, and perform node classification using traditional ML methods, may bypass {\poisonedgnn}. The codes for these platforms are not released. Hence we could not confirm their resilience to {\poisonedgnn} by experiments.}

 \blue{One way to extend {\poisonedgnn} to node-wise HT classification is by training another GNN, called the payload model. The payload model accepts the circuits and looks for the backdoor trigger. If the backdoor trigger is detected, the payload model can influence the node-wise classification as non-Trojan, similar to the approach presented in~\cite{response_reference_y}.}

%% file: texfiles/Sec6_Conclusion.tex
\section{Conclusion}
\label{sec:conclusion}

In this work, we developed {\poisonedgnn} the first backdoor attack against GNNs processing digital circuits. 
An adversary can embed a crafted sub-circuit (backdoor trigger) to some training circuits and alter their labels to an attacker-chosen target label. 
As a result, the GNN trained on the backdoored dataset is forced to associate the backdoor trigger with the target label beneficial to the attacker. 
To demonstrate this alarming vulnerability of GNNs, we consider two case studies of (i)~hiding hardware Trojans (HTs) and (ii)~intellectual property (IP) piracy. 
Through {\poisonedgnn}, we show that backdoored GNNs can misclassify HT-infected circuits as HT-free and misclassify pirated designs as IP-piracy-free, with an attack success rate of $100\%$. 
This work sheds light on the vulnerabilities of GNNs and the importance of shielding the GNNs when employed in security-critical applications such as HT or piracy detection.

%% file: bios.tex
\begin{IEEEbiography}[{\includegraphics[width=1in,height=1.25in,clip,keepaspectratio]{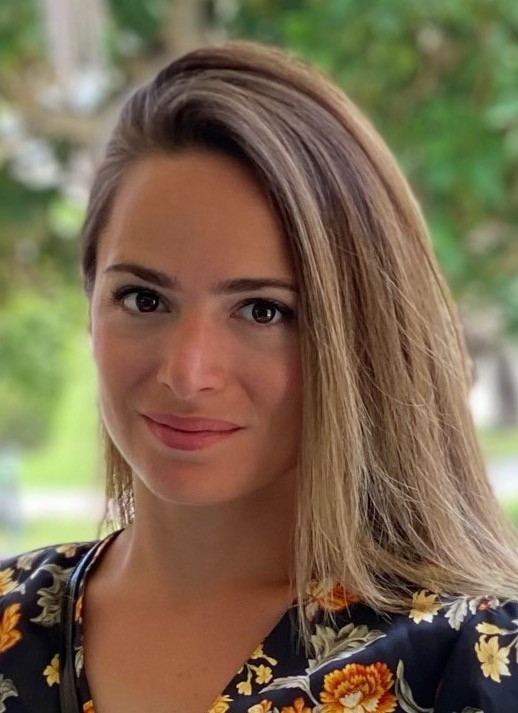}}]{Lilas~Alrahis} is a Postdoctoral Associate at New York University Abu Dhabi and an International Karlsruhe Institute of Technology (KIT) excellence fellow. She received the M.Sc.\ degree and the Ph.D.\ degree in electrical and computer engineering from Khalifa University, UAE, in 2016 and 2021, respectively. Her research interests include Hardware Security, Design for Trust, Logic Locking, and Applied Machine Learning. 
Dr.\ Alrahis won the MWSCAS Myril B.\ Reed Best Paper Award in 2016 and the Best Paper Award at the Applied Research Competition held in conjunction with Cyber Security Awareness Week, in 2019. Dr. Alrahis is currently serving as Associate Editor of the Integration, the VLSI Journal. 
\end{IEEEbiography}

\begin{IEEEbiography}[{\includegraphics[width=1in,height=1.25in,clip,keepaspectratio]
{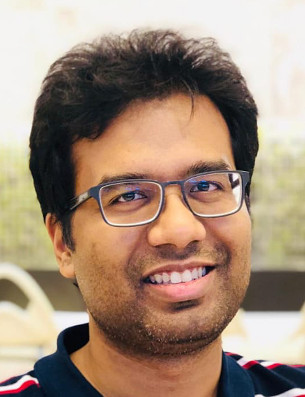}}]{Satwik~Patnaik} is a Postdoctoral Researcher with the Department of Electrical and Computer Engineering, Texas A\&M University, College Station, TX, USA.
He received his Ph.D.\ degree in Electrical engineering from Tandon School of Engineering, New York University, NY, USA, in September 2020.
His research delves into IP protection techniques, CAD frameworks for incorporating security, leveraging the 3D paradigm for security, exploiting security properties of emerging devices, and utilizing machine learning and reinforcement learning techniques for enhancing hardware security.
Dr.\ Patnaik received the Bronze Medal in the Graduate Category at the ACM/SIGDA Student Research Competition held at ICCAD 2018, the Best Paper Award at the Applied Research Competition (ARC) held in conjunction with Cyber Security Awareness Week (CSAW) in 2017, and the third place at ARC in 2021.
He is currently co-organizing a first-of-its-kind hardware security competition, AI vs. Humans co-located with CSAW 2022.
\end{IEEEbiography}

\begin{IEEEbiography}[{\includegraphics[width=1in,height=1.25in,clip,keepaspectratio]
{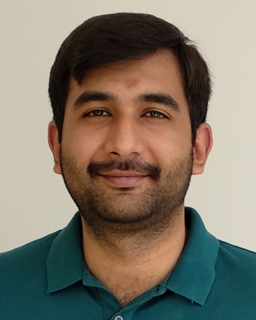}}]{Muhammad~Abdullah~Hanif} received the M.Sc. degree in electrical engineering from National University of Sciences and Technology, Pakistan. He was a University Assistant with the Department of Informatics, Institute of Computer Engineering, Technische Universit{\"a}t Wien (TU~Wien), Austria. He is currently pursuing the Ph.D. degree in computer engineering from TU~Wien and working at New York University Abu~Dhabi, UAE. His current research interests include brain-inspired computing, machine learning, approximate computing, computer architecture, energy-efficient design, robust computing, system-on-chip design, and emerging technologies.
\end{IEEEbiography}

\begin{IEEEbiography}[{\includegraphics[width=1in,height=1.25in,clip,keepaspectratio]{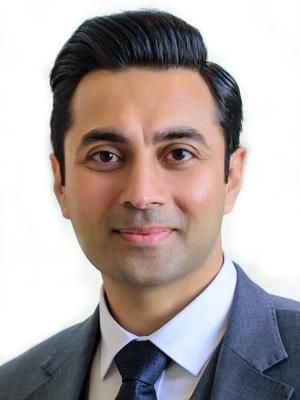}}]{Muhammad Shafique}
received the Ph.D.\ degree in computer science from the Karlsruhe Institute of Technology (KIT), Germany, in 2011.
In 2016, he joined the Institute of Computer Engineering at the Faculty of Informatics, Technische Universit{\"a}t Wien, Austria as a Professor of Computer Architecture and Robust, Energy-Efficient Technologies. 
Since 2020, he is with New York University Abu Dhabi, UAE.
His research interests include brain-inspired computing, AI, \& ML hardware and system-level design, energy-efficient systems, robust computing, hardware security, emerging technologies, FPGAs, MPSoCs, and embedded systems. 

Dr.\ Shafique holds one U.S.\ patent has (co-)authored 6 Books, 10+ Book Chapters, and over 200 conference and journal papers. He received the 2015 ACM/SIGDA Outstanding New Faculty Award, AI 2000 Chip Technology Most Influential Scholar Award in 2020 and 2022, six gold medals, and several best paper awards.
\end{IEEEbiography}

\begin{IEEEbiography}[{\includegraphics[width=1in,height=1.25in,clip,keepaspectratio]{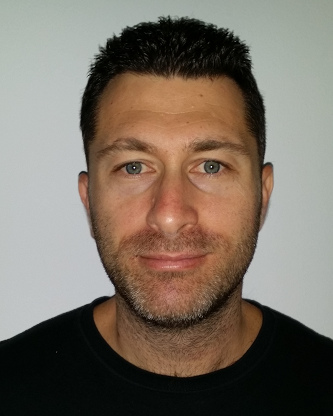}}]{Ozgur Sinanoglu} is a professor of electrical and computer engineering at New York University Abu Dhabi. 
He obtained his Ph.D.\ in Computer Science and Engineering from University of California San Diego. 
He has industry experience at TI, IBM and Qualcomm.
During his Ph.D.\ he won the IBM Ph.D.\ fellowship award twice. 
He is also the recipient of the best paper awards at IEEE VLSI Test Symposium 2011 and ACM Conference on Computer and Communication Security 2013. 

Prof.\ Sinanoglu’s research interests include design-for-test, design-for-security and design-for-trust for VLSI circuits, where he has more than 200 conference and journal papers, and 20 issued and pending US Patents. 
Prof.\ Sinanoglu is the director of the Center for CyberSecurity at NYU Abu Dhabi. 
His recent research is being funded by US National Science Foundation, US Department of Defense, Semiconductor Research Corporation, Intel Corp, and Mubadala Technology.
\end{IEEEbiography}